\documentclass[fleqn,usenatbib]{mnras}

\usepackage{newtxtext,newtxmath}
\usepackage[T1]{fontenc}

\DeclareRobustCommand{\VAN}[3]{#2}
\let\VANthebibliography\thebibliography
\def\thebibliography{\DeclareRobustCommand{\VAN}[3]{##3}\VANthebibliography}

\usepackage{hyperref}
\hypersetup{colorlinks=true,allcolors=blue,pdfborder={0 0 0}}
\usepackage{graphicx}	% Including figure files
\title[Stirred but not shaken: HD 16743's debris disc]{Stirred but not shaken: a multi-wavelength view of HD~16743's debris disc}

\author[J. P. Marshall et al.]{
Jonathan P. Marshall,$^{1,2}$\thanks{E-mail: jmarshall@asiaa.sinica.edu.tw (JPM)}
J. Milli,$^{3}$
E. Choquet,$^{4}$
C. del Burgo,$^{5}$
G. M. Kennedy,$^{6,7}$
F. Kemper,$^{8,9,10}$
\newauthor
M. C. Wyatt,$^{11}$
Q. Kral,$^{12}$
R. Soummer$^{13}$\\
$^{1}$Academia Sinica Institute of Astronomy and Astrophysics, 11F of AS/NTU Astronomy-Mathematics Building, No.1, Sect. 4, Roosevelt Rd, Taipei 10617, Taiwan\\
$^{2}$Centre for Astrophysics, University of Southern Queensland, Toowoomba, QLD 4350, Australia\\
$^{3}$Universit\'e Grenoble Alpes, CNRS, IPAG, 38000 Grenoble, France\\
$^{4}$Aix Marseille Univ, CNRS, CNES, LAM, Marseille, France\\
$^{5}$Instituto Nacional de Astrof\'isica \'Optica y Electr\'onica, Luis Enrique Erro \#1, CP 72840, Tonantzintla, Puebla, M\'exico\\
$^{6}$Department of Physics, University of Warwick, Gibbet Hill Road, Coventry, CV4 7AL, UK\\
$^{7}$Centre for Exoplanets and Habitability, University of Warwick, Gibbet Hill Road, Coventry CV4 7AL, UK\\
$^{8}$Institut de Ciencies de l'Espai (ICE, CSIC), Can Magrans, s/n, 08193 Bellaterra, Barcelona, Spain\\
$^{9}$ICREA, Pg. Llu\'is Companys 23, Barcelona, Spain\\
$^{10}$Institut d'Estudis Espacials de Catalunya (IEEC), E-08034 Barcelona, Spain\\
$^{11}$Institute of Astronomy, University of Cambridge, Madingley Road, Cambridge CB3 0HA, UK\\
$^{12}$LESIA, Observatoire de Paris, Universit\'e PSL, CNRS, Sorbonne Universit\'e, Univ. Paris Diderot, Sorbonne Paris Cit\'e, 5 place Jules Janssen, 92195 Meudon, France\\
$^{13}$Space Telescope Science Institute, Baltimore, MD 21218, USA\\
}
\date{Accepted XXX. Received YYY; in original form ZZZ}

\pubyear{2023}

\begin{document}
\label{firstpage}
\pagerange{\pageref{firstpage}--\pageref{lastpage}}
\maketitle

\begin{abstract}
Planetesimals -- asteroids and comets -- are the building blocks of planets in protoplanetary discs and the source of dust, ice and gas in debris discs. Along with planets they comprise the left-over material after star formation that constitutes a planetary system. Planets influence the dynamics of planetesimals, sculpting the orbits of debris belts to produce asymmetries or gaps. We can constrain the architecture of planetary systems, and infer the presence of unseen planetary companions, by high spatial resolution imaging of debris discs. HD~16743 is a relatively young F-type star that hosts a bright edge-on debris disc. Based on far-infrared \textit{Herschel} observations its disc was thought to be stirred by a planetary companion. Here we present the first spatially resolved observations at near-infrared and millimetre wavelengths with \textit{HST} and ALMA, revealing the disc to be highly inclined at $87\fdg3~^{+1\fdg9}_{-2\fdg5}$ with a radial extent of 157.7$^{+2.6}_{-1.5}$~au and a FWHM of 79.4$^{+8.1}_{-7.8}$~au ($\Delta R/R = 0.5$). The vertical scale height of the disc is $0.13~\pm~0.02$, significantly greater than typically assumed unstirred value of 0.05, and could be indicative of stirring of the dust-producing planetesimals within the disc by bodies at least a few times the mass of Pluto up to 18.3~$M_{\oplus}$ in the single object limit.  

\end{abstract}


%%%%%%%%%%%%%%%%%%%%%%%%%%%%%%%%%%%%%%%%%%%%%%%%%%

%%%%%%%%%%%%%%%%% BODY OF PAPER %%%%%%%%%%%%%%%%%%

\section{Introduction}

Planetesimals, asteroidal or cometary bodies grown from the agglomeration of dust grains or pebbles, are believed to form rapidly in belts within protoplanetary discs \citep{2021Lovell}. These bodies are the material that fuels the assembly of planets, either gas giants which grow within the lifetime of the gas-rich protoplanetary disc, in the first few Myr of the star's life \citep{1996Pollack,2022SuarezMascareno}, or rocky planets that may form much later, perhaps up to 100~Myr \citep{1998Kokubo,2001Chambers,2004Chambers,2023Johansen}. Studies of the largest and brightest protoplanetary discs \citep{2018Andrews} revealed annular structures thought to be indicative of the gravitational influence of nascent proto-planets on the surrounding disc \citep{2018Dong,2018Huang,2018Zhang}, due to their lack of correlation with the temperature structure \citep{2018Long,2019vanderMarel}. Those large, bright protoplanetary discs are thought to be analogous to the largest, brightest debris discs we observe around much older stars \citep{2021vanderMarel,2021Michel,2022Najita}. The inheritance of structure from the protoplanetary to debris disc phase is still uncertain \citep{2019Lodato}, as is the mass budget required to create the planetary systems we have observed \citep{2018Manara,2020Tychoniec,2021Mulders,2021KrivovWyatt}.

Debris discs around main sequence stars are tenuous, dusty structures generated by collisions within those remnant planetesimal belts that survived both the dispersal of their protoplanetary disc and the simultaneous and/or subsequent planet formation processes \citep[e.g.][]{2018Hughes}. In contrast to their progenitor protoplanetary discs they are generally gas-poor, although increasing evidence suggests that substantial masses of CO gas may reside in the youngest and more massive debris disc systems, predominantly around A- and F-type stars \citep{2016Greaves,2017Moor,2017Kral,2020Marino,2022Hales}, but more recently also later type stars \citep{2019bMatra,2020Kral}. Detailed characterisation of these systems requires multi-wavelength, spatially resolved imaging to determine both the spatial distribution of the dust-producing planetesimals, and the optical properties of the dust grains \citep[e.g.][]{2014Marshall}. At far-infrared and millimetre wavelengths observations trace thermal emission from the largest and coldest dust grains in the debris disc, a proxy for the location of the planetesimals \citep[e.g.][]{2018Matra,2021Marshall}. At near-infrared wavelengths observations measure the scattered light from small grains revealing their albedo, shape, and scattering properties \citep[e.g.][]{2014Schneider,2020Esposito}.

We can identify the influence of unseen planetary companions on debris discs with high resolution imaging observations at millimetre wavelengths to trace the shape and structure of the planetesimal belts \citep[e.g.][]{2021Marino,2021Faramaz}. Theoretical expectations place constraints on the timescales required for the dynamical stirring of debris discs, either by planetesimals within the belt \citep{2018KrivovBooth}, or by an external perturber (planet) \citep{2009MustillWyatt}. Young, bright debris discs are especially suited for this work. Previously, the influence of planetary companions has been inferred from analysis of bright debris discs with \textit{Herschel} \citep{2015Moor,2016Vican}, but the modest angular resolution of that facility, combined with its operational wavelength in the far-infrared lead to some ambiguity in the interpretation of the structures observed for those discs, e.g. the width of the debris disc was rarely resolved \citep[e.g.][]{2021Marshall}. 

At millimetre wavelengths with interferometric facilities (e.g. ALMA, SMA), much higher angular resolution observations are possible. Imaging observations with ALMA have revealed structures inferred to be the result of disc-planet interactions \citep{2022Pearce}, such as eccentric debris discs \citep{2017Macgregor,2019Faramaz}, gaps within broad debris discs \citep{2018Marino,2019Marino}, two populations of planetesimals required to explain the vertical distribution of dust around $\beta$ Pictoris \citep{2019bMatra}, and the halo of millimetre dust grains around several stars, including HR~8799 \citep{2018Macgregor,2019Geiler}. Studies of debris discs radii vs. brightness revealed a strong decline in brightness in the first 100~Myr, suggested to be the result of embedded planets depleting their host discs \citep{2021Pawellek}. Furthermore, that same work identified that the debris discs around young F-type stars in the $\beta$ Pictoris moving group were narrower than expected, again inferred to be the influence of planetary companions on the planetesimal belts. A separate work found that debris discs with narrow belts were more likely to be eccentric \citep{2020Kennedy}, consistent with the notion that a planetary companion was perturbing the belt (although the eccentricity could equally be imprinted in the protoplanetary disc phase).

In this work we present a detailed investigation of the architecture and dust properties of the debris disc around HD~16743, a young, F-type star estimated to be between 10 and 50~Myr old \citep{2011Moor}, and previously identified as potentially being stirred by a planetary companion \citep{2015Moor}. We combine spatially resolved imaging in near-infrared scattered light and millimetre wavelength thermal emission with ancillary photometry and spectroscopy to characterise the system. The rest of the paper is laid out as follows: in Section \ref{sec:observations} we present details of the new ALMA and VLT/SPHERE observations, along with existing archival data. The modelling approaches to determine the disc architecture and dust properties are then summarised along with their outcomes in Section \ref{sec:results}. We give context to these results in Section \ref{sec:discussion} through comparison with similar debris discs. Finally, we give a summary of our findings and our conclusions in Section \ref{sec:conclusions}.

\section{Observations}
\label{sec:observations} 

In this section we present various data sets including ALMA, \textit{Herschel}, \textit{HST}/NICMOS and VLT/SPHERE imaging data, \textit{Spitzer}/IRS mid-infrared spectroscopy, and optical to mid-infrared photometry from Tycho, 2MASS, \textit{WISE}, and \textit{Akari}. In combination, these measurements span the stellar photosphere and disc continuum and scattered light contributions to the total observed emission. We present the reduction process for the ALMA, \textit{HST}/NICMOS, and VLT/SPHERE observations in detail, whilst the remaining data were taken from public archives and catalogues.

\subsection{ALMA}

ALMA Band 6 observations of HD~16743 were taken in Cycle 6 as part of project 2019.1.01220.S (PI: J.P. Marshall). These data were obtained from the ESO ALMA Science Archive\footnote{\href{http://almascience.eso.org/aq/}{http://almascience.eso.org/aq/}}. The target was observed over two scheduling blocks on 17th and 19th December 2019 achieving close to the requested continuum sensitivity (12~$\mu$Jy achieved cf. 10~$\mu$Jy requested, 1.6 hrs total on-source time), with an angular resolution of $\simeq 1\arcsec$ (baselines from 15 to 313 m). The spectral setup consisted of four windows; three windows were set up to measure the continuum, each with 128 channels over 2~GHz bandwidth. A fourth window covered the $^{12}$CO (2-1) line at 230.538 GHz and sampled its 1.5~GHz bandwidth with 1916 channels (1.278 kms$^{-1}$). In combination, the four channels provided a total of 7.5~GHz bandwidth to study the target emission. 

Calibration and reduction of the ALMA observation were carried out in CASA 5.6 using scripts provided by the observatory. Image reconstruction was carried out using the \textit{tclean} task combining all four spectral windows for the greatest signal-to-noise after visual inspection of the baseband covering the CO (2-1) molecular line revealed no significant emission. We reconstruct the image using Briggs weighting with a robustness parameter of 0.5 so as to maximise the sensitivity to faint emission whilst retaining the angular resolution. The Briggs-weighted continuum image used in the analysis presented here has an r.m.s. noise of 15~$\mu$Jy/beam. The dirty beam has an ellipsoidal FWHM $1\farcs25~\times~0\farcs98$ at a position angle of 70$\fdg$8, equivalent to a spatial resolution of 36$~\times~$28 au.

The disc is detected in continuum emission with a peak signal-to-noise ratio in excess of 10 at the disc ansae, and signal-to-noise ratio greater than 5 across the whole disc extent. The disc is oriented roughly NNW-SSE with a near edge-on presentation, highly inclined compared to predictions based on prior \textit{Herschel} imaging. The continuum image and visibilities are shown in Figure \ref{fig:hd16743_alma_cont}. We do not find any significant emission associated with the CO (2-1) line, which is shown in Figure \ref{fig:hd16743_alma_line}.

\subsection{VLT/SPHERE}

HD~16743 (HIP~12361; Gaia DR3 4742097275828451584) was observed with the high-contrast imager VLT/SPHERE \citep{Beuzit2019} as part of the SPHERE High-Angular Resolution Debris Disks Survey\footnote{ESO programs 096.C-0388 and 097.C-0394} \citep[SHARDDS,][]{Wahhaj2016,Choquet2017,2018Marshall,Cronin-Coltsmann2021}. This survey is an imaging search aimed at resolving and characterising new debris discs never detected in scattered light around stars within 100 pc, having an infrared excess greater than $10^{-4}$. It uses the IRDIS subsystem \citep{Dohlen2008} in broad band H ($\lambda=1.625\,\mu$m, $\Delta\lambda=0.290\,\mu$m) and the apodised Lyot coronagraph of diameter 185 mas. HD~16743 was observed on the night of October 3rd 2015. The coronagraphic observations lasted $\sim40$ minutes. They were carried out in pupil-stabilised mode to allow the use of Angular Differential Imaging \citep[ADI,][]{Marois2006} to subtract the stellar halo. The star was observed symmetrically about meridian, which led to a total parallactic angle rotation of $21\degr$, or 22 resolution elements at $2\farcs5$. In total, the sequence gathered 288 coronagraphic images of individual Detector Integration Time (DIT) 8s. The DIMM seeing ranged between 0\farcs75 and 1\arcsec\, and the achieved Strehl was $\sim80\%$.

To mitigate the problem of self-subtraction of any astrophysical signal inherent to ADI \citep[e.g.][]{Milli2012}, Reference Differential Imaging \citep[RDI,][]{Ruane2019} was also applied as an alternative data reduction technique to subtract the glare of the central star. 

The disc is marginally detected at the ansae, as shown in Figure \ref{fig:hd16743_vlt-sphere}. The disc geometry is consistent with the ALMA image. It has a mean surface brightness of 40~$\mu$Jy.arcsec$^{-2}$ in $H$-band, but has been strongly affected by the image processing.

\subsection{HST/NICMOS}

We acquired \textit{HST}/NICMOS observations of HD~16743 from the Hubble archive. These observations were originally taken as part of program GO-11157 (PI: J. Rhee), a search for debris discs around 22 targets with strong \textit{IRAS} infrared excess. HD~16743 was observed on July 23 2007 with the coronagraphic imaging mode of the NIC2 camera ($0\farcs07565$~pixel$^{-1}$, focal plane mask radius 0\farcs3). The observations were obtained in the two wide-band filters, F110W ($\lambda=1.115~\mu$m, $\Delta\lambda=0.562~\mu$m 95\%-integrated bandwidth) and F160W ($\lambda=1.601~\mu$m, $\Delta\lambda=0.390~\mu$m 95~\%-integrated bandwidth). Each filterband image was taken with five frames at two different spacecraft orientations (30$\degr$ apart) to reduce the impact of unattenuated PSF artefacts on the recovery of faint extended structure. Total on-source integration time was 38.4 mins in the F110W filter and 39.5 mins in the F160W filter. 

These data were reduced and combined (de-rotated, stacked) using an advanced version of the pipeline developed for the ALICE program (PI: R. Soummer), a consistent reanalysis of the \textit{HST}/NICMOS coronagraphic archives with advanced starlight subtraction methods \citep{Choquet2014,Hagan2018}, which allowed the discovery of 12 other new debris discs in scattered light \citep{Soummer2014, Choquet2016, Choquet2017,2018Choquet,2018Marshall}. To subtract the stellar contribution in each images of both data sets and reveal the faint disc around HD~16743, we used large libraries of reference star images assembled from the NICMOS coronagraphic archives in the corresponding bandpass, down-selected to match the raw image of the target by cross-correlation (F110W library: 281 frames from 39 different reference stars; F160W library: 397 frames from 38 different reference stars). In these libraries, we also added the images of the target obtained in the complementary telescope roll, which provide the best PSF matches and significantly helped recovering the disc signal,  albeit producing some self-subtraction near the star (limited effect thanks to the edge-on geometry of the disk). We computed the stellar PSF models with the PCA-KLIP algorithm \citep{Soummer2012} using 33\% of the principal components of the libraries, after masking the central part of the image within a radius of 14 pixels. These models were then subtracted from each target image to reveal the faint disk, and the images were derotated to North up, mean-combined, and scaled to Jy.arcsec$^{-2}$ using the NICMOS photometric calibration values.

The disc is detected with a mean signal-to-noise ratio on the disc surface of 4.2 per pixel in the F110W image and 2.9 per pixel in the F160W image. The disc has a geometry consistent with the ALMA image. It has a mean surface brightness of 53~$\mu$Jy.arcsec$^{-2}$ in the F110W filter, and 40~$\mu$Jy.arcsec$^{-2}$ in the F160W filter. The images are shown in Figure \ref{fig:hd16743_hst-nicmos} (left).

\subsection{Ancillary data}

We complement the ALMA, \textit{HST}/NICMOS, and VLT/SPHERE observations described above with a range of spectroscopic and photometric data taken from the literature. These data include optical and near-infrared photometry from Gaia, Tycho, and 2MASS \citep{2016Gaia,2018Gaia,2000Hog,2006Skrutskie}, mid-infrared photometry from \textit{Akari}, \textit{WISE} and \textit{Spitzer} \citep{2010Ishihara,2010Wright,2011Moor}, the \textit{Spitzer}/IRS spectrum taken from the Combined Atlas of Sources with \textit{Spitzer} IRS Spectra\footnote{https://cassis.sirtf.com/atlas/} \citep[CASSIS; ][]{2011Lebouteiller}, and the \textit{Herschel}/PACS and /SPIRE far-infrared and sub-millimetre imaging observations taken from the \textit{Herschel} Science Archive\footnote{http://archives.esac.esa.int/hsa/whsa/} as level 2.5 pipeline reduced, mosaicked data products. A summary of the photometry used in the radiative transfer modelling is provided in Table \ref{tab:photometry}.

\begin{table}
	\centering
	\caption{Photometry used in the radiative transfer modelling.}
	\label{tab:photometry}
	\begin{tabular}{cccc}
    	\hline\hline
		Wavelength & Flux Density & Telescope / &  Reference \\
		 ($\mu$m)  &    (mJy)     & Instrument  &            \\
    	\hline
        0.44 & 5625~$\pm$~51 & Johnson $B$ & H\o g 2000 \\
        0.55 & 7047~$\pm$~66 & Johnson $V$ & H\o g 2000 \\
        0.64 & 6808~$\pm$~62 & \textit{Gaia} $G$ & Gaia DR3 \\
        0.78 & 7146~$\pm$~66 & Cousins $I$ & Wu 2013 \\
        1.24 & 5544~$\pm$~132 & 2MASS $J$ & Skrutskie 2006 \\
        1.65 & 4296~$\pm$~102 & 2MASS $H$ & Skrutskie 2006\\
        2.16 & 2851~$\pm$~73 & 2MASS $K_{\rm s}$ & Skrutskie 2006\\
        3.4 & 1236~$\pm$~154 & \textit{WISE} W1 & Wright 2010 \\
        4.6 & 770~$\pm$~41 & \textit{WISE} W2 & Wright 2010 \\
        9 & 267~$\pm$~9 & \textit{Akari} IRC9 & Ishihara 2010 \\
        12 & 130~$\pm$~6 & \textit{WISE} W3 & Wright 2010 \\
        22 & 56~$\pm$~3 & \textit{WISE} W4 & Wright 2010 \\
        24 & 50~$\pm$~2 & \textit{Spitzer} MIPS & Mo\'or 2011 \\
        70 & 388~$\pm$~26 & \textit{Spitzer} MIPS & Mo\'or 2011 \\
        100 & 369~$\pm$~27 & \textit{Herschel} PACS & Mo\'or 2015 \\
        160 & 174~$\pm$~24 & \textit{Spitzer} MIPS & Mo\'or 2011 \\
        160 & 226~$\pm$~32 & \textit{Herschel} PACS & Mo\'or 2015\\
        250 & 82~$\pm$~6 & \textit{Herschel} SPIRE & Mo\'or 2015 \\
        350 & 38~$\pm$~6 & \textit{Herschel} SPIRE & Mo\'or 2015 \\
        1270 & 1.235~$\pm$~0.131 & ALMA Band 6 & This work \\
    	\hline
    \end{tabular}
\end{table}

\section{Modelling and Results}
\label{sec:results}

Here we present the process of our analysis. We begin by determining the extent and orientation of the disc at millimetre wavelengths, searching for any evidence of disc-planet interaction in the revealed architecture, and line emission from molecular CO gas. We then use the measured disc architecture as a constraint in the radiative transfer modelling to determine the minimum size of dust grains in the disc, the size distribution of these grains, and their total mass. Thereafter, we combine the continuum emission modelling with scattered light observations to determine the scattering albedo of the dust grains.

\begin{figure*}
	\includegraphics[width=\columnwidth]{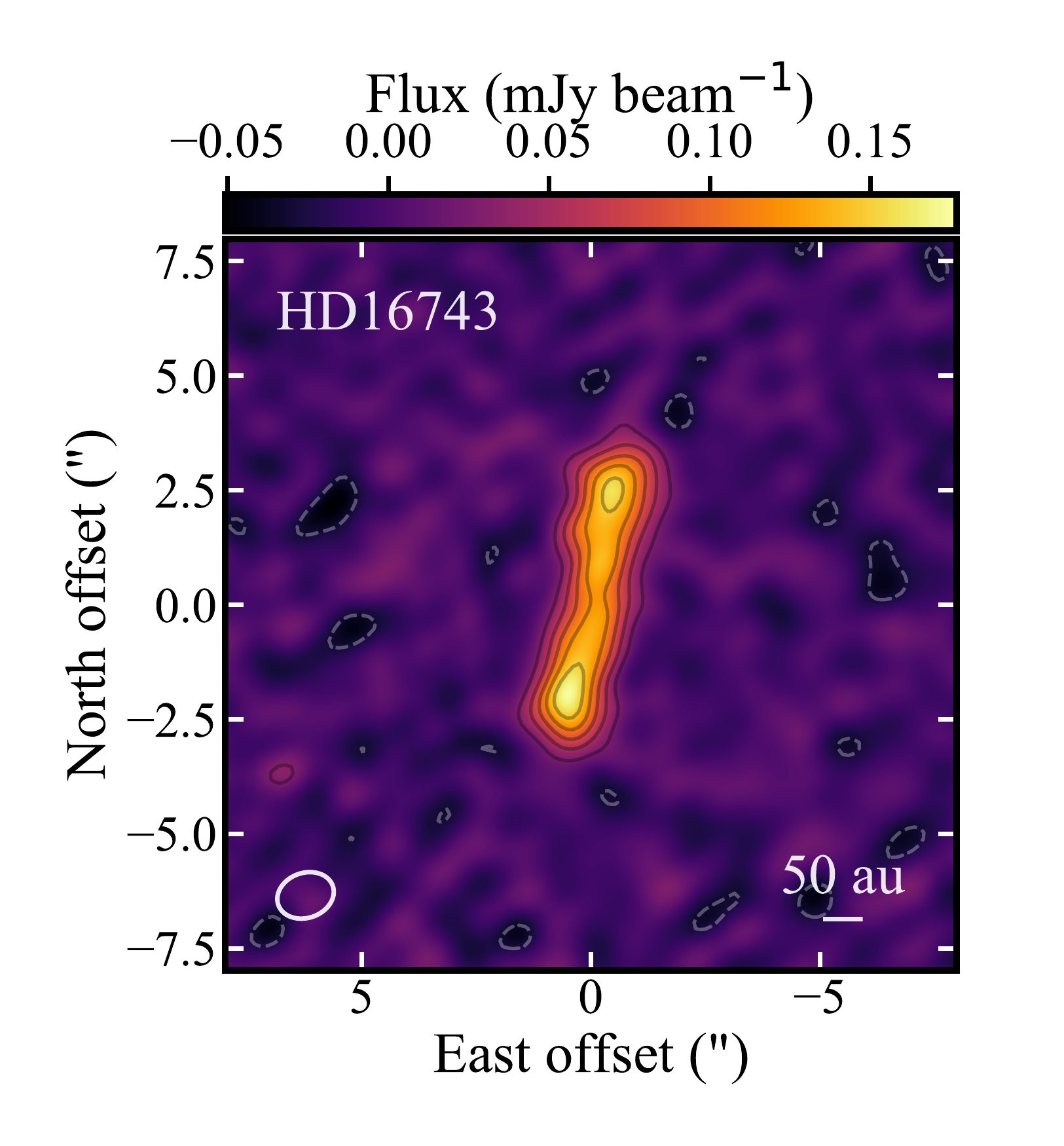}
	\includegraphics[width=\columnwidth]{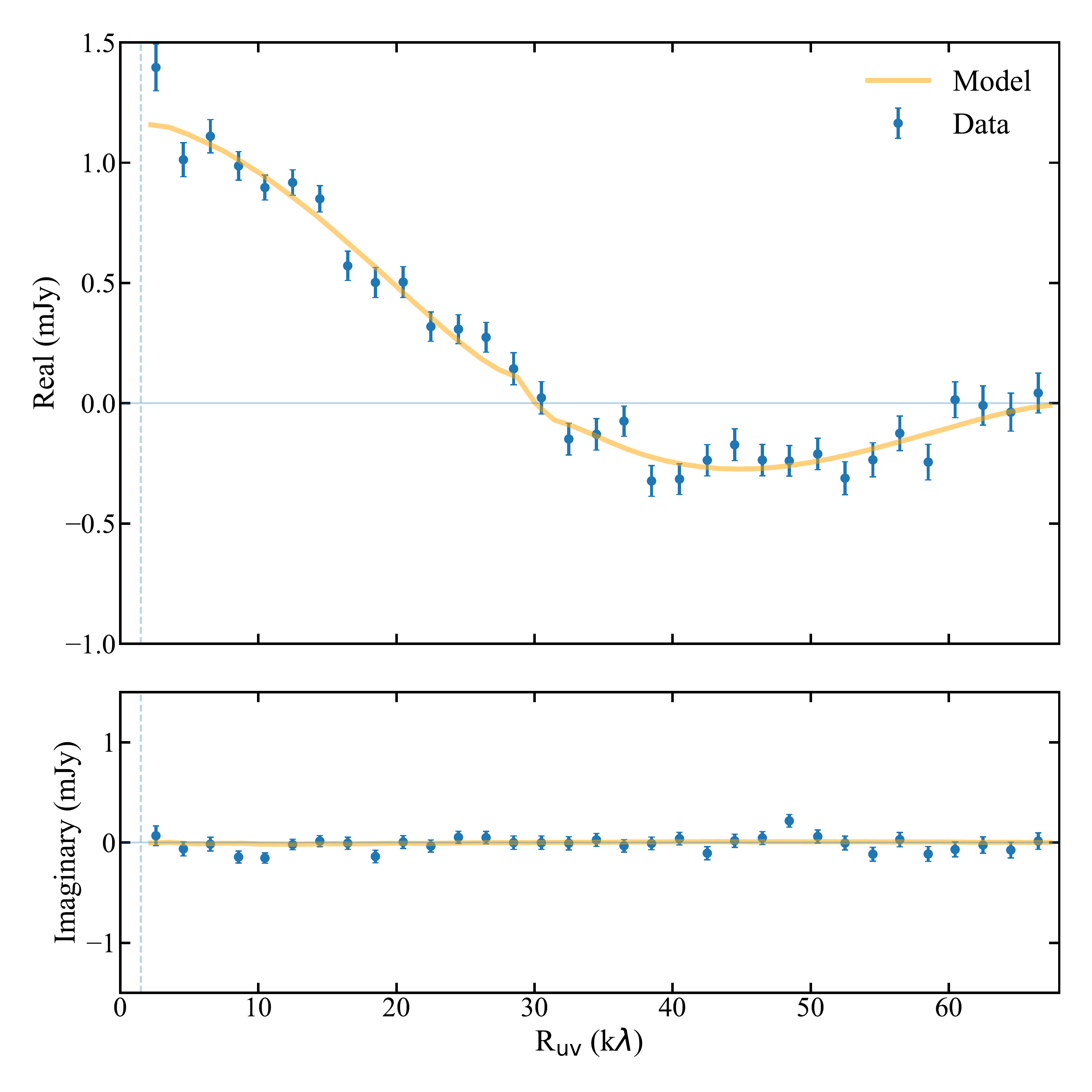}
    \caption{\textit{Left}: ALMA Band 6 continuum image of HD~16743. The image has been \textit{clean}ed and reconstructed with a Briggs weight of 0.5. There is no evidence for a star-disc offset, or a second component to the disc, from fitting the visibilities. The instrument beam ($0\farcs95~\times~0\farcs67$, $\phi = 88~\degr$) is denoted by the white ellipse in the bottom left corner. Contours are in steps of 2-$\sigma$ from $\pm$2-sigma, with broken contours denoting negative values. Orientation is north up, east left. \textit{Right}: Visibility modelling using a single Gaussian annulus to fit to the data points, no significant residuals indicative of a second disc component, or structure to the disc, are present after model subtraction.}
    \label{fig:hd16743_alma_cont}
\end{figure*}

\subsection{Millimetre emission}

We model the disc as a single Gaussian annulus. Its architecture is defined by a flux density $f_{\rm disc}$, peak radius $R_{\rm peak}$ and full width at half maximum $R_{\rm fwhm}$, scale height $h = z/R$, inclination $i$, and position angle $\phi$. Additionally, we include parameters for the stellar photospheric contribution $f_{\star}$, and position offsets $\Delta$RA, $\Delta$Dec between the star (assumed to be at the phase centre of the observations) and the disc centre. Previous spatially resolved \textit{Herschel} observations of the disc at far-infrared wavelengths provide good constraints on the radius and inclination \citep{2015Moor}, although the disc appears much more inclined to the line-of-sight in the ALMA millimetre imaging.

We use the Python-based code Modelling Interferometric Array Observations\footnote{https://github.com/dlmatra/miao} ({\sc Miao}, Luca Matr\`a) to compare disc models to the observed image in the visibility plane. For clarity, a brief summary of the modelling process using {\sc Miao} is given here. The $u$,$v$ spacings, real and imaginary visibilities, and their weights are first calculated by combining and averaging the channels of the four spectral windows of the calibrated measurement set into a single channel using the CASA task \textit{mstransform}, reducing the size of the data set considerably in the process. A model image of the inclined disc surface brightness profile is generated using {\sc radmc-3d} \citep{2012Dullemond}, rotated according to the required position angle, and the star is added as a point source shifted relative to the image centre according to the star-disc offset. This model is then convolved with the primary beam of the interferometer. Finally, synthetic visibilities are generated by taking the Fourier transform of model images evaluated at the same $u$,$v$ points as the observation using the {\sc Galario} package \citep{2018Tazzari}. The least squares sum of the weighted synthetic visibilities is then calculated to determine the quality of the model's representation of the observations. To explore the parameter space and determine the maximum probability model, we use the package {\sc emcee} \citep{2013ForemanMackey}.  The ensemble sampler is set up using ten walkers per free parameter in the model (nine parameters) and 1\,000 steps for a total of 90\,000 realisations. The first 800 steps of the run are discarded as burn-in, leaving the posterior probability distribution to be evaluated from the remaining 18\,000 realisations of the model. The walkers are initialised at values derived from a 2D Gaussian fit to the source brightness profile in the ALMA image (for $R_{\rm peak}$, $R_{\rm fwhm}$, $i$, and $\phi$), a scale height of 0.03, flux densities of 1~mJy for the disc and 0.1~mJy for the star, and an assumed zero offset for the stellar position. The walkers are then given an additional random uniform scatter of up to $\pm$~10~\% of those values to create their initial starting positions. 

The results of the disc fitting are presented in Table \ref{tab:hd16743_alma_fit}, along with the ALMA continuum image and $u$,$v$ plot for the observations and maximum probability model in Figure \ref{fig:hd16743_alma_cont}. The disc extent $R_{\rm peak} = 157.7^{+2.6}_{-1.5}$~au, width $R_{\rm fwhm} = 79.4^{+8.1}_{-7.8}$~au, and position angle $\phi = 168$\fdg$5^{+0.6}_{-0.5}$, are consistent with the interpretation of the \textit{Herschel}/PACS far-infrared imaging observations \citep{2015Moor,2021Marshall}. The inclination of the system is close to edge-on, with an inclination $i = 87$\fdg$3^{+1.9}_{-2.5}$, this is much more inclined than the orientation expected based on the \textit{Herschel}/PACS observations (around $58\fdg5~\pm~8\fdg3$). HD~16743 was only marginally resolved along the major axis of the disc in those data, so the revision to a steeper inclination is not so drastic, and possibly related to variation observed in the PACS PSF \citep{2012Kennedy}. Based on the absence of significant residuals ($\geq$~3-$\sigma$) after the subtraction of the maximum likelihood model from the observation, there is no evidence for non-axisymmetric structure to the disc indicative of stirring by a companion. This is consistent with the recent collisional modelling of debris discs that interpreted the extent and brightness of the HD~16743 system as being consistent with self-stirring \citep{2018KrivovBooth}. We also find no evidence in the modelling for an additional disc component to the system, despite the presence of substantial mid-infrared excess in the SED, although this may be attributed to the sensitivity of the ALMA observations being too low to detect the warm component. Likewise, the stellar photosphere is also not detected at the sensitivity of the ALMA observations, and there is no significant star-disc offset inferred from the modelling. 

\begin{table}
	\centering
	\caption{Results from visibility fits to the ALMA observation.}
	\label{tab:hd16743_alma_fit}
	\begin{tabular}{lc}
    	\hline\hline
		Parameter & Value \\
    	\hline
    	$R_{\rm peak}$ (au) & 157.7$^{+2.6}_{-1.5}$ \\
    	$R_{\rm fwhm}$ (au) & 79.4$^{+8.1}_{-7.8}$ \\
    	$h$ & 0.131$^{+0.014}_{-0.016}$ \\
    	$i$ (\degr) &  87.3$^{+1.9}_{-2.5}$ \\
    	$\phi$ (\degr) & 168.5$^{+0.6}_{-0.5}$ \\
    	$f_{\rm dust}$ (mJy) & 1.235~$\pm$~0.131 \\
    	$f_{\rm star}$ ($\mu$Jy) & 18.9$^{+12.3}_{-9.9}$ \\
    	$\Delta$RA (\arcsec) & -0.05~$\pm$~0.02 \\
        $\Delta$Dec (\arcsec) & 0.03~$\pm$~0.04 \\
    	\hline
    \end{tabular}
\end{table}

\begin{figure}
    \centering
    \includegraphics[width=\columnwidth,trim={0cm 0cm 0cm 1cm},clip]{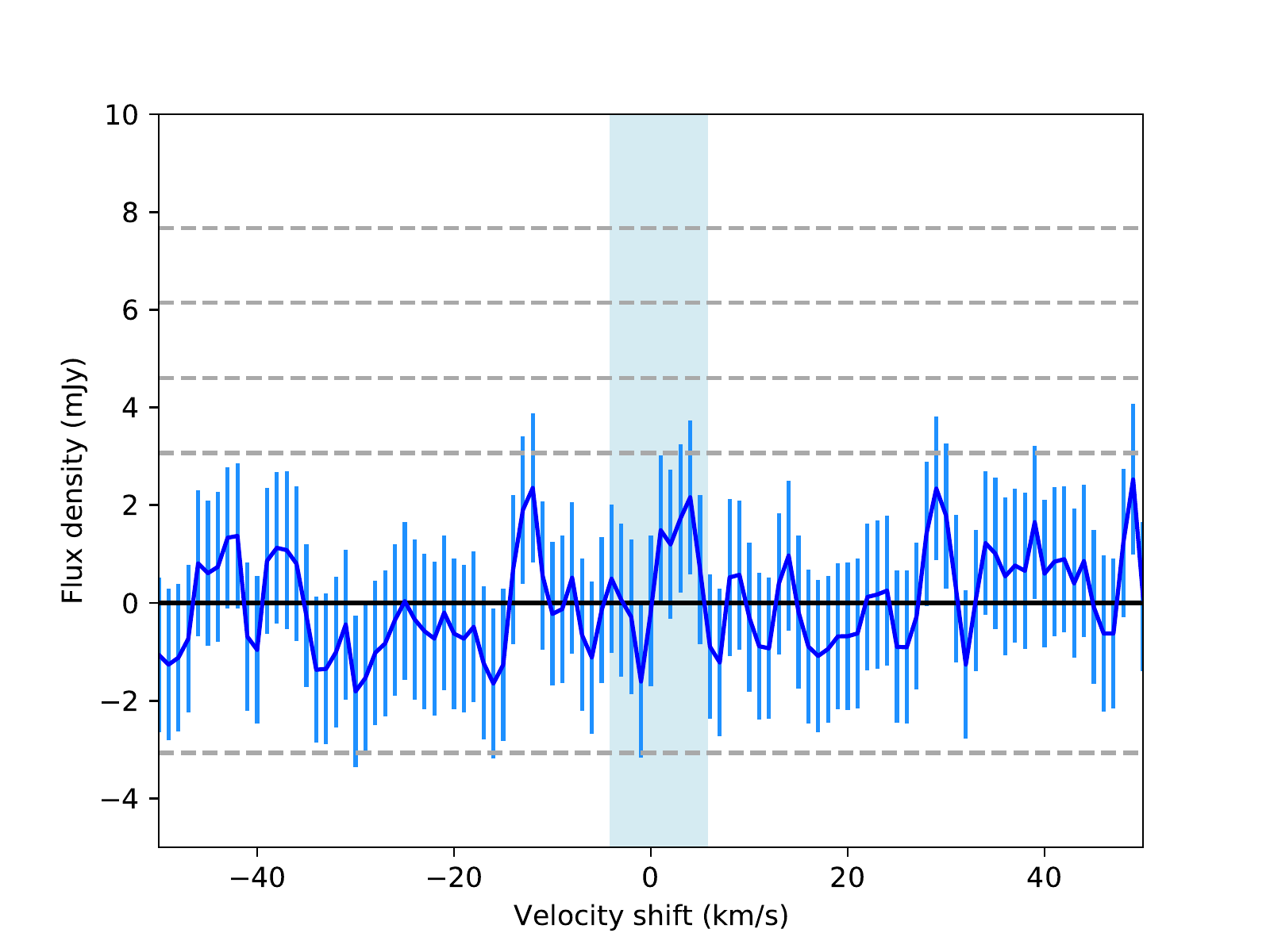}
    \caption{ALMA CO spectrum centred on a rest frequency of 230.538~GHz. The blue shaded region denotes the 10 km/s wide channel used to determine the presence of CO emission, centred on the stellar velocity. Uncertainties are 1-$\sigma$. The grey dashed lines denote uncertainties from $\pm$2$\sigma$ in steps of 1-$\sigma$.}
    \label{fig:hd16743_alma_line}
\end{figure}

Having identified the extent and orientation of the disc in the continuum image, we then use that to search for gas emission from the disc in the spectral window covering the CO (2-1) line at 230.538 GHz. We define a spatial mask for the spectral window consisting of pixels which have $\geq 3$-$\sigma$ emission in the continuum image. A spectrum is then extracted for each pixel in that mask and an independent frequency shift is applied to each spectrum based on the projected semi-major axes of the pixel centres and the Keplerian velocity at that distance from the star. We then interpolate the individual spectra to a set of common velocities and sum over the pixel spectra to produce a final spectrum for CO emission from the system , which is presented in Figure \ref{fig:hd16743_alma_line}. We find no evidence for any CO emission associated with the disc, and obtain a 5-$\sigma$ upper limit of 17.1 mJy/beam in a 10 km/s wide channel, equivalent to 1.3$\times10^{-22}$~W/m$^{2}$, from the observation. This constraint is far above the predicted CO emission level of 1.4$\times10^{-24}$~W/m$^{2}$ ($M_{\rm CO}$ = 2.8$\times10^{-7}~M_{\oplus}$) from \citet{2017Kral}.

\subsection{Fundamental stellar parameters}
\label{sec:star_params}
\begin{table}
    \centering
    \caption{Fundamental stellar parameters of HD~16743. The absolute magnitude M$_G$, obtained from the G apparent magnitude and the parallax $\Pi$ of \textit{Gaia} DR3, the colour BP-RP from the same data release, and the age estimate of \citet{Desidera2015}, were held as fixed values to infer the fundamental parameters from the Bayesian code of \citet{2016delBurgo,2018delBurgo}. 
    %We also adopted the prior that the star is on the PMS. 
    \label{tab:stellar_parameters}}
    \begin{tabular}{lcc}
    \hline\hline 
        Parameter & Value & Ref. \\
    \hline
        $\Pi$ ($mas$) & 17.299 $\pm$ 0.018 & Gaia DR3 \\   
        \hline
        M$_G$ (mag) & 2.896 $\pm$ 0.004 & Gaia DR3 \\       
        BP-RP (mag) & 0.478 $\pm$ 0.005 & Gaia DR3 \\  
        Age (Ma)   & 100$^{+400}_{-60}$ & Desidera et al.\\
        \hline
        Effective temperature (K) & 6953$^{+29}_{-14}$ & this work\\
        Radius ($R_{\odot}$) & 1.535$^{+0.012}_{-0.021}$ & this work\\
        Mass ($M_{\odot}$) & 1.535$^{+0.05}_{-0.03}$ & this work\\
        Surface gravity ($\log g$/$cgs$) & 4.252$^{+0.027}_{-0.017}$ & this work\\
        Luminosity ($\log L$/$L_{\odot}$) & 0.695$^{+0.005}_{-0.006}$ & this work\\
        Bolometric magnitude (mag) & 3.002$^{+0.015}_{-0.012}$ & this work\\
        {[Fe/H]}             & 0.24$^{+0.20}_{-0.17}$ & this work\\
    \hline
    \end{tabular}
\end{table}

 %The PMS hypothesis is supported by the following. 
As found in the \textit{Gaia} Data Release 3 \citep[DR3, ][]{2021Gaia}, HD~16743 has a parallax $\Pi$= 17.299$\pm$0.018 $mas$, a proper motion specified by $\mu_{\rm RA}$= 73.546$\pm$0.017 $mas~yr^{-1}$ and $\mu_{\rm DEC}$= 49.560$\pm$0.020 $mas~yr^{-1}$), and a radial velocity v$_R$= 14.55$\pm$0.40 km s$^{-1}$. These values are likened to those of the close pair formed by HD~16699 ($\Pi$= 17.273$\pm$0.016~$mas$; $\mu_{\rm RA}$= 74.749$\pm$0.016~$mas~yr^{-1}$ and $\mu_{\rm DEC}$= 48.728$\pm$0.017~$mas~yr^{-1}$; v$_R$= 16.14$\pm$0.18 km s$^{-1}$) and HD~16699B (SAO~232842; $\Pi$= 16.69$\pm$0.21~$mas$; $\mu_{\rm RA}$= 68.39$\pm$0.20~$mas~yr^{-1}$ and $\mu_{\rm DEC}$= 49.79$\pm$0.23~$mas~yr^{-1}$; v$_R$= 15.43$\pm$1.15 km s$^{-1}$). Given these stars likely form a dynamically linked system, it is wise to assume a common origin for them.

The latter statement was already heeded by \citet{2011Moor}, who supported the youth of HD~16743 based on the age indicators of SAO~232842. They emphasised that this star preserves a high lithium (Li) abundance and that it is an X-ray source with $L_{\rm X}/L_{\rm bol} = -3.32$, a ratio comparable with those of similar spectral type stars in the Pleiades and in young nearby moving groups. Yet, \citet{Desidera2015} found that SAO~232842 is a close visual binary with a separation between components of 0.06 arcsec (3.6 au) and ${\Delta}H$=0.2$\pm$0.2 mag from VLT/NaCo observations, and pointed out that the age indicators, including the Li equivalent width of 250 m\AA, are puzzling. They methodically examined all relevant information before determining an age of 100$^{+400}_{-60}$ Ma. This roughly corresponds to the age of the Pleiades, bounded by those of the younger IC 4665 and the older Hyades open star clusters \citep{2022Cantat-Gaudin}.

In order to infer the fundamental stellar parameters of HD 16743, we have employed the Bayesian inference code of \citet{2016delBurgo,2018delBurgo} applied to the PARSEC v1.2S library of stellar evolution models \citep{2012Bressan}. Its statistical comparison with dynamical masses of detached eclipsing binaries proves that the combined method and library are an apposite choice, especially for main-sequence stars, where the predicted masses are, on average, within 4~\% of accuracy \citep{2018delBurgo}. 

We run the code in a similar way to \citet{2016delBurgo}, but taking as inputs the absolute $G$ magnitude, $M_G$, and the colour $BP-RP$ from \textit{Gaia} DR3, and the age propounded by \citet{Desidera2015}. We assumed the star is not affected by interstellar extinction, which is justified by the small distance of 57.81$\pm$0.06~pc. Conversely to \citet{2018delBurgo}, we did not arrive at a solution hooked to the most likely evolution phase, but permitted the code to wield all models (particularly, those of the pre-main and main sequence) compatible with the inputs and their uncertainties. Table \ref{tab:stellar_parameters} displays these inputs and the inferred fundamental stellar parameters of HD~16743.

As a sanity check, we applied the same code on SAO~232842, under the premise that it is composed of two stars alike. This is supported by the work of \citet{Desidera2015}, who noticed that the components are indistinguishable by the ASAS photometry. The assertion allowed us to deduce a righteous first-order approximation for the photometric inputs of each star from the unresolved \textit{Gaia} data. Here, we adopted, as a prior, that they are on the pre-main-sequence and have an iron to hydrogen abundance ratio [Fe/H] = 0.2$\pm$0.2, i.e., slightly above but still compatible with Solar metallicity. The latter served as an input parameter, along with the estimated absolute G magnitude and the BP-RP colour from \textit{Gaia} DR3. The resulting age of 57$\pm$19 Ma differs from that (namely, 16$\pm$3 Ma) obtained if we ignore that SAO~232842 is a binary star, while it is consistent with the findings of \citet{Desidera2015}. Specifically, our lower limit of the age matches theirs.

\subsection{Spectral energy distribution}

We model HD~16743's spectral energy distribution (SED) as a star plus two dust components - a modified blackbody to represent the warm excess present at mid-infrared wavelengths, and a parametric model to represent the cold excess at mid-infrared to millimetre wavelengths.

The stellar contribution to the SED is fitted by interpolation between stellar photosphere models from the BT-NEXTGEN grid \citep{2006Barber,2009Asplund,2011Allard,2012Allard}. The best-fit model is determined by least-squares fitting of synthetic photometry from the calculated models to the observations at wavelengths between 0.4 and 10~$\mu$m. We adopt values for the stellar parameters as determined in the previous section, i.e. $T_{\star} = 6953$~K, $R_{\star} = 1.535~R_{\odot}$, and $L_{\star} = 4.95~L_{\odot}$, we also assume Solar metallicity, and use the distance of 57.806~pc derived from the \textit{Gaia} DR3 parallax \citep{2021Gaia}.

The warm modified blackbody component of the disc is defined by its temperature $T_{\rm warm}$, break wavelength $\lambda_{\rm 0}$, and sub-millimetre slope $\beta$. These parameters are determined by simultaneously fitting two modified blackbody components to the photometry with $\lambda > 10~\mu$m and s/n $> 3$. It is assumed that both modified blackbody components have the same break wavelength and slope, so the $\lambda_{\rm 0}$ and $\beta$ of the warm component are dictated by the fit to the cold component. This assumption is not necessarily true, due to size-dependent radial migration in the disc impacting the size distribution at a given radial distance, but the current observations do not directly probe the sub-millimetre spectral slope of the warm component. The best-fit parameters for the two components are determined by least-squares fitting to the observations. We find that the inner warm component has a best-fit temperature $T_{\rm warm} = 120~$K, with $\lambda_{\rm 0} = 200~\mu$m and $\beta = 1.5$ (based on the cold component), and a fractional luminosity ($L_{\rm dust}/L_{\star}$) of $7\times10^{-5}$. The warm component fitted in this way is subtracted from the observed photometry before fitting the parameteric model to the cold component so as not to bias the results of that fit.

\begin{figure}
	\includegraphics[width=\columnwidth]{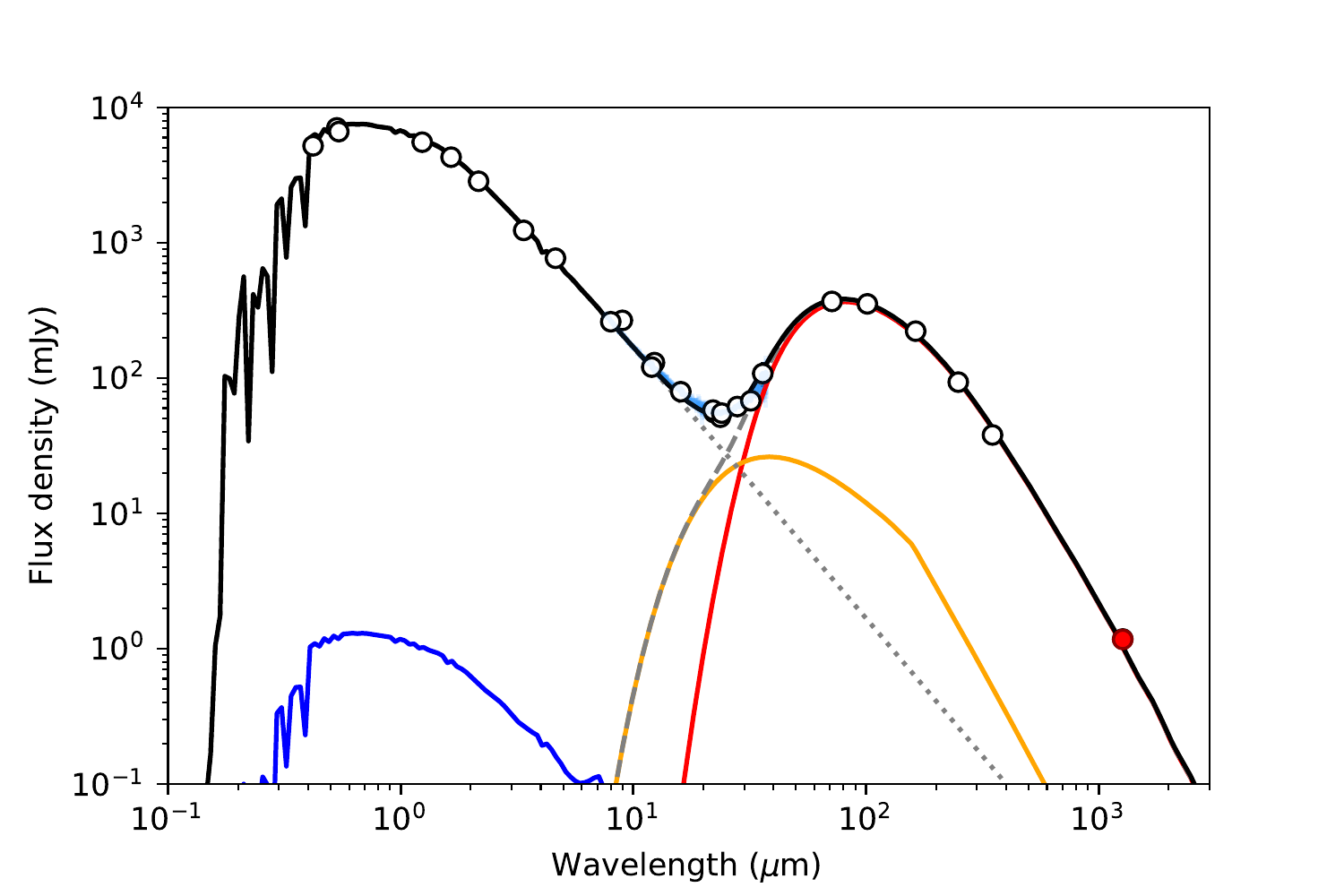}
    \caption{Spectral energy distribution of HD~16743. White data points are ancillary photometry from the literature, whilst the red data point denotes the new ALMA observation. The light blue line denotes the \textit{Spitzer}/IRS spectrum. The stellar photosphere model is denoted by the grey dotted line, the disc emission by the grey dashed line, and the total emission by the black solid line. The blue, orange, and red solid lines denote the disc scattered light, warm, and cold components, respectively.}
    \label{fig:hd16743_sed}
\end{figure}

The cold component of the disc is defined by the radial extent and width, $R_{\rm peak}$ and $R_{\rm fwhm}$ (inferred from the ALMA observations), the minimum and maximum size of the dust grains, $a_{\rm min}$ and $a_{\rm max}$ (fixed as 1~mm), the exponent of the dust power law size distribution $q$, the total mass of dust $M_{\rm dust}$, and its composition \citep[assumed to be astronomical silicates,][]{2003Draine}. From the modified blackbody fit, the cold component has a fractional luminosity ($L_{\rm dust}/L_{\star}$) of $3.2\times10^{-4}$. We use the Python MCMC package {\sc emcee} \citep{2013ForemanMackey} to determine the maximum probability parameters for the cold component model. We use 30 walkers (10 per parameter) and 500 steps to explore the parameter space with {\sc emcee}, initialising the walkers at values of 3.0~$\mu$m for $a_{\min}$, 3.5 for $q$, and $10^{-4}~M_{\oplus}$ for $M_{\rm dust}$, plus a uniform random scatter of 10~\% for each value. For each set of parameters a radiative transfer model of the system is run. Synthetic photometry is then calculated from that model at the relevant wavelengths for comparison with the observations and their associated uncertainties. The posterior probability distribution was constructed from the final 100 steps of the chains (3\,000 realisations of the model). The maximum probability and its uncertainties are determined using the 16th, 50th, and 84th percentiles of the posterior probability distribution. A summary of the model parameters, their ranges, and the results of the fitting are given in Table \ref{tab:hd16743_sed_fit}, whilst the best-fit model SED and the observations are presented in Figure \ref{fig:hd16743_sed}.

\begin{table}
	\centering
	\caption{Results of radiative transfer modelling.}
	\label{tab:hd16743_sed_fit}
	\begin{tabular}{lcc}
    	\hline\hline
		Parameter & Range & Value \\
    	\hline
    	Composition & --- & astron. sil. \\
		$a_{\rm min}$ ($\mu$m) & 0.5 -- 50.0 & 5.01$^{+0.11}_{-0.09}$\\
		$a_{\rm max}$ ($\mu$m) & --- & 1\,000 \\
		$q$ & 3 -- 4 & 3.72~$\pm$~0.03 \\
		$M_{\rm dust}$ ($\times10^{-3}~M_{\oplus}$) & 0.01 -- 10.0 & 4.38~$\pm$~0.06 \\
		$\chi^{2}$ & n/a &  45.7 \\
		\hline
	\end{tabular}
\end{table}

\subsection{Scattered light}

\begin{figure}
	\includegraphics[width=0.48\textwidth]{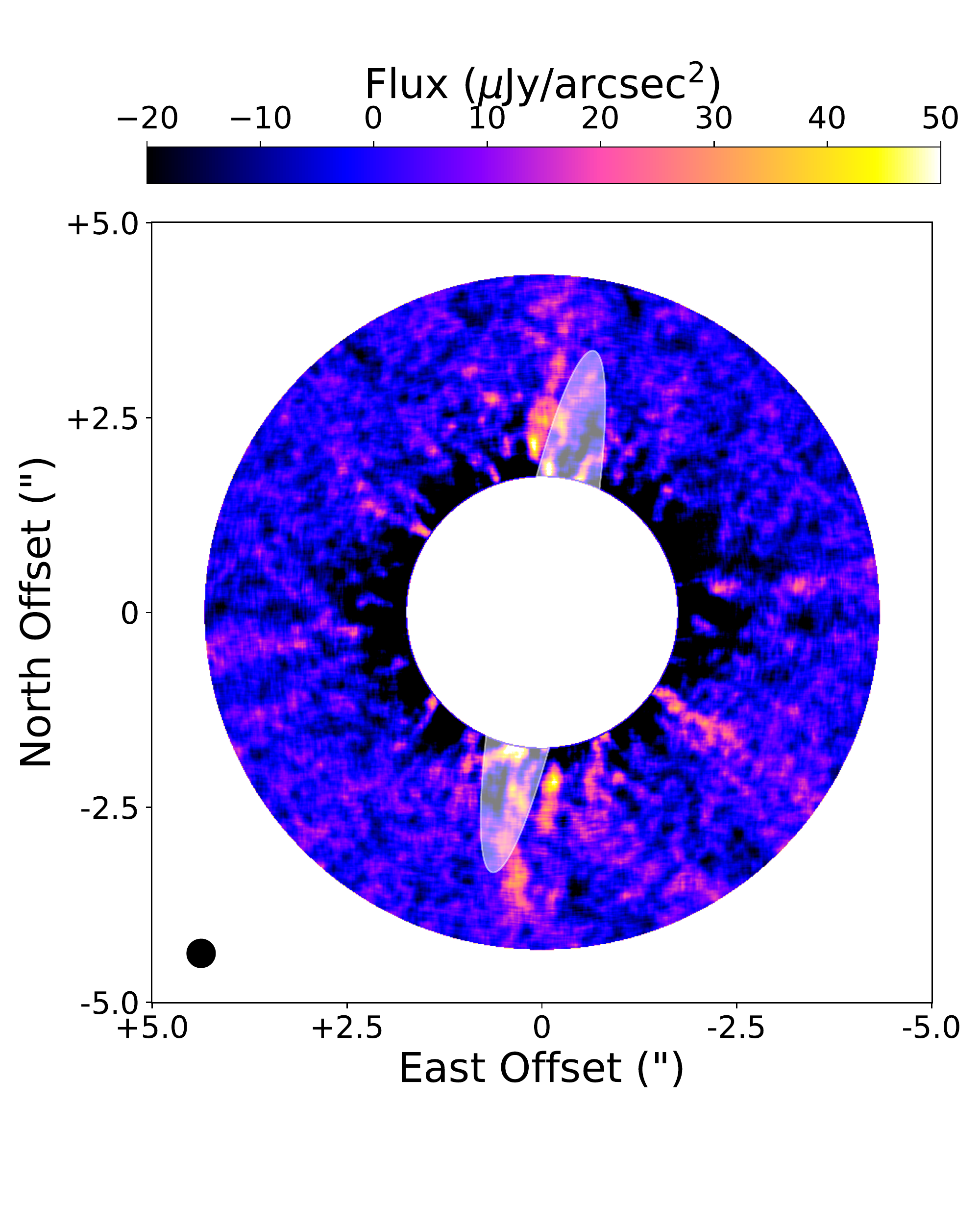}
	\vspace{-0.25in}
    \caption{VLT/SPHERE $H$-band radial ADI filtered image of HD~16743. The shaded ellipse denotes the orientation and extent of the debris disc from the ALMA image. The central 1\farcs7 radius region of the image is dominated by residuals from the starlight suppression and has been masked in this plot to highlight fainter structure relevant to tracing the debris disc. Orientation is north up, east left. The instrument beam FWHM is denoted by the black circle in the lower left corner of the image.}
    \label{fig:hd16743_vlt-sphere}
\end{figure}

\begin{figure*}
	\includegraphics[width=0.33\textwidth]{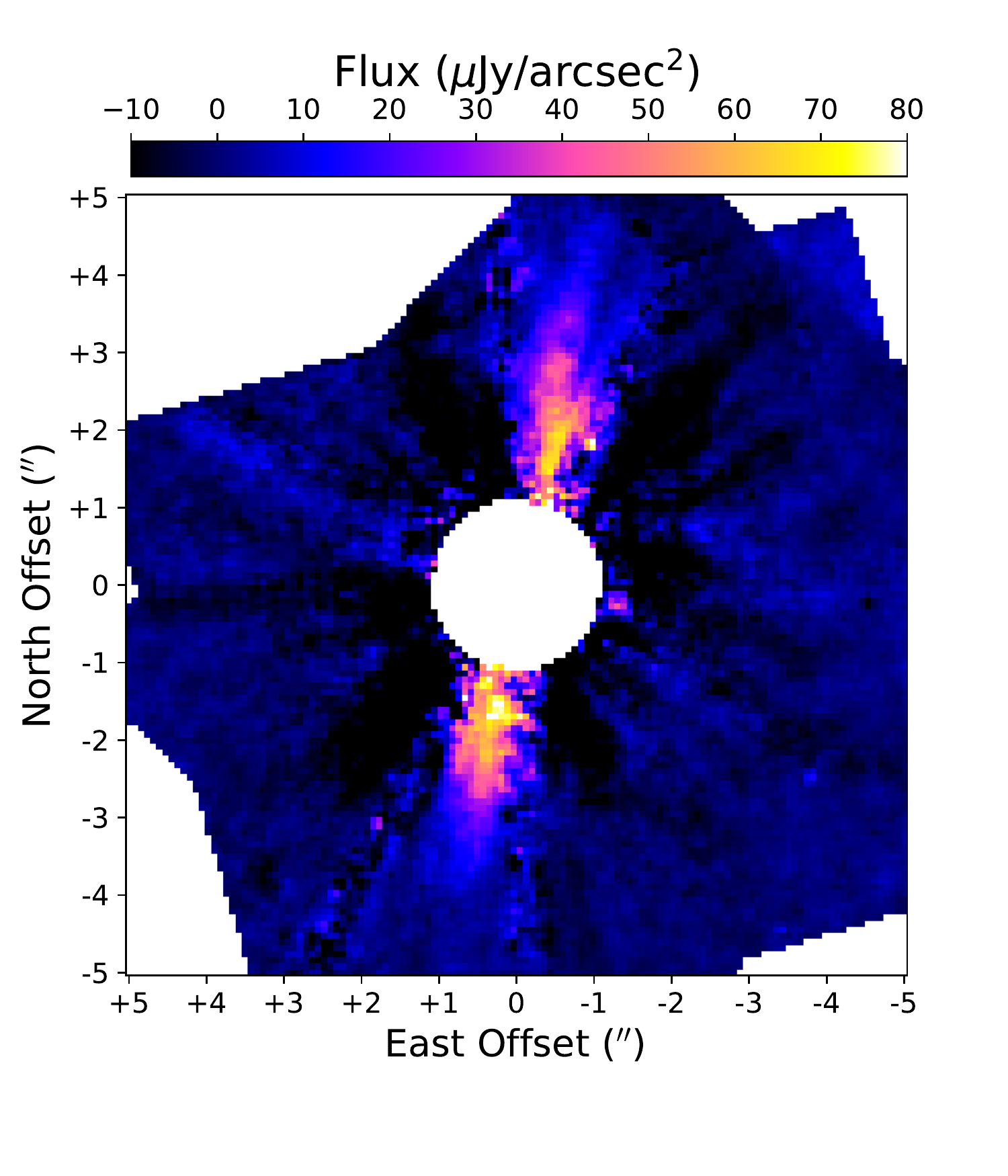}
	\includegraphics[width=0.33\textwidth]{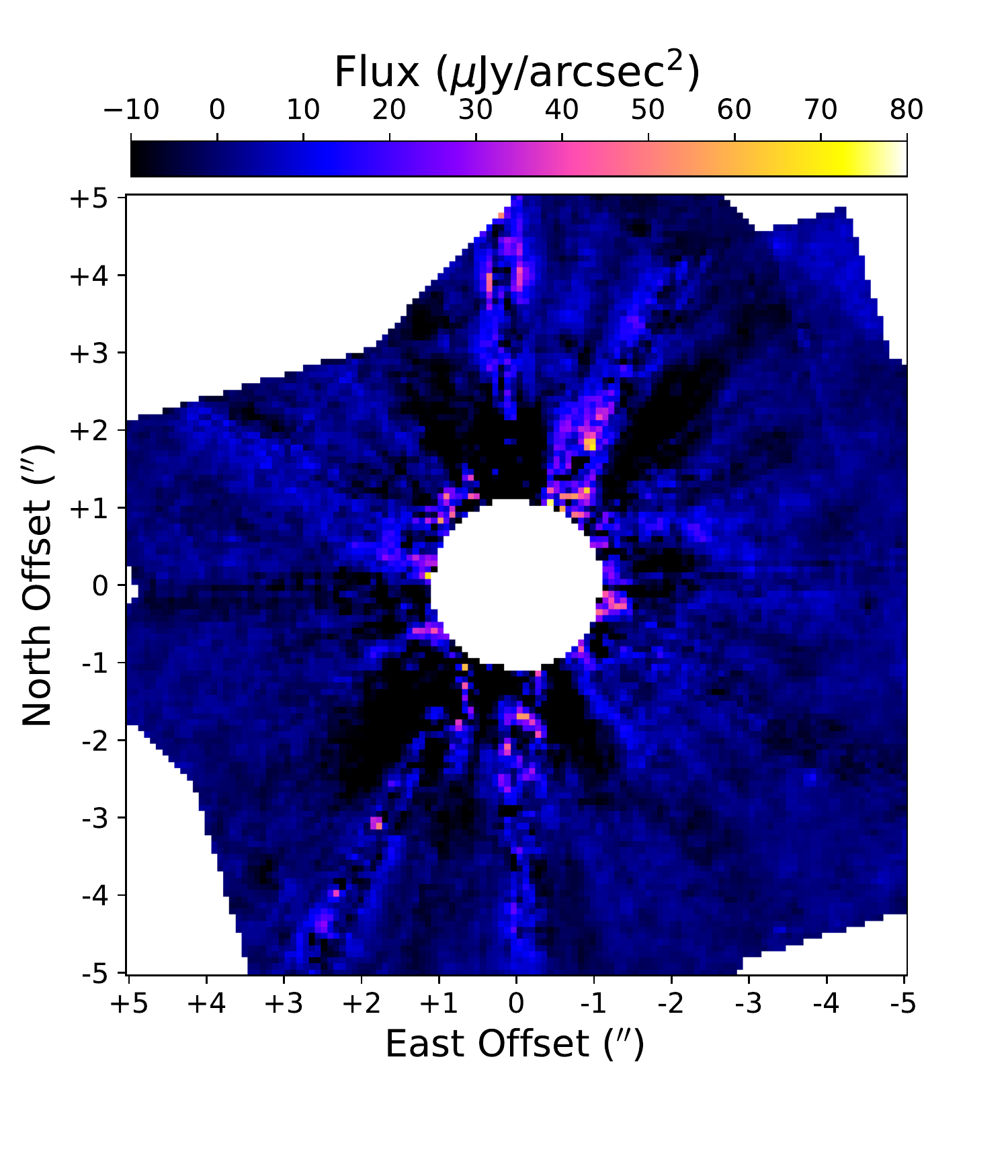}
	\includegraphics[width=0.33\textwidth]{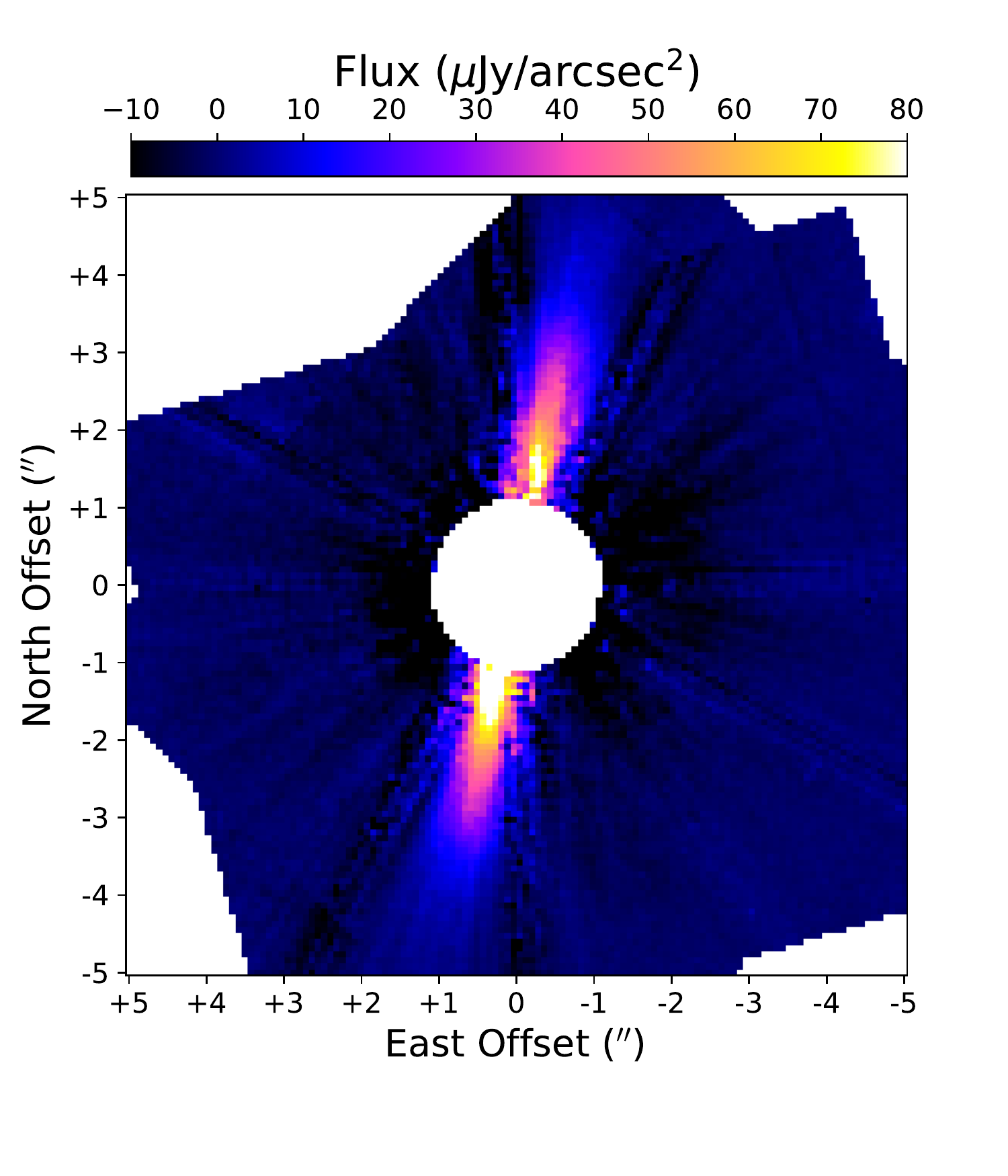}
	\\
	\includegraphics[width=0.33\textwidth]{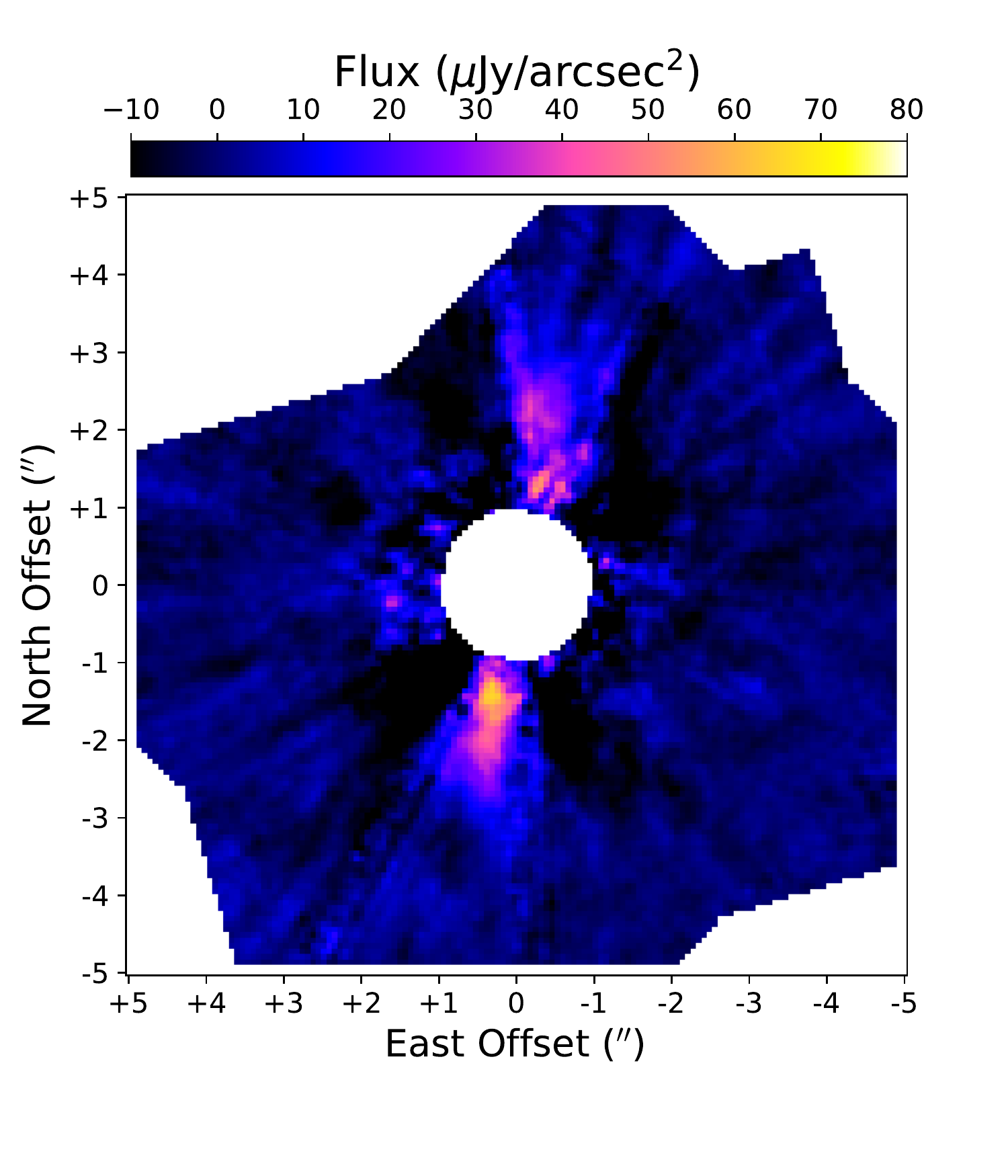}
	\includegraphics[width=0.33\textwidth]{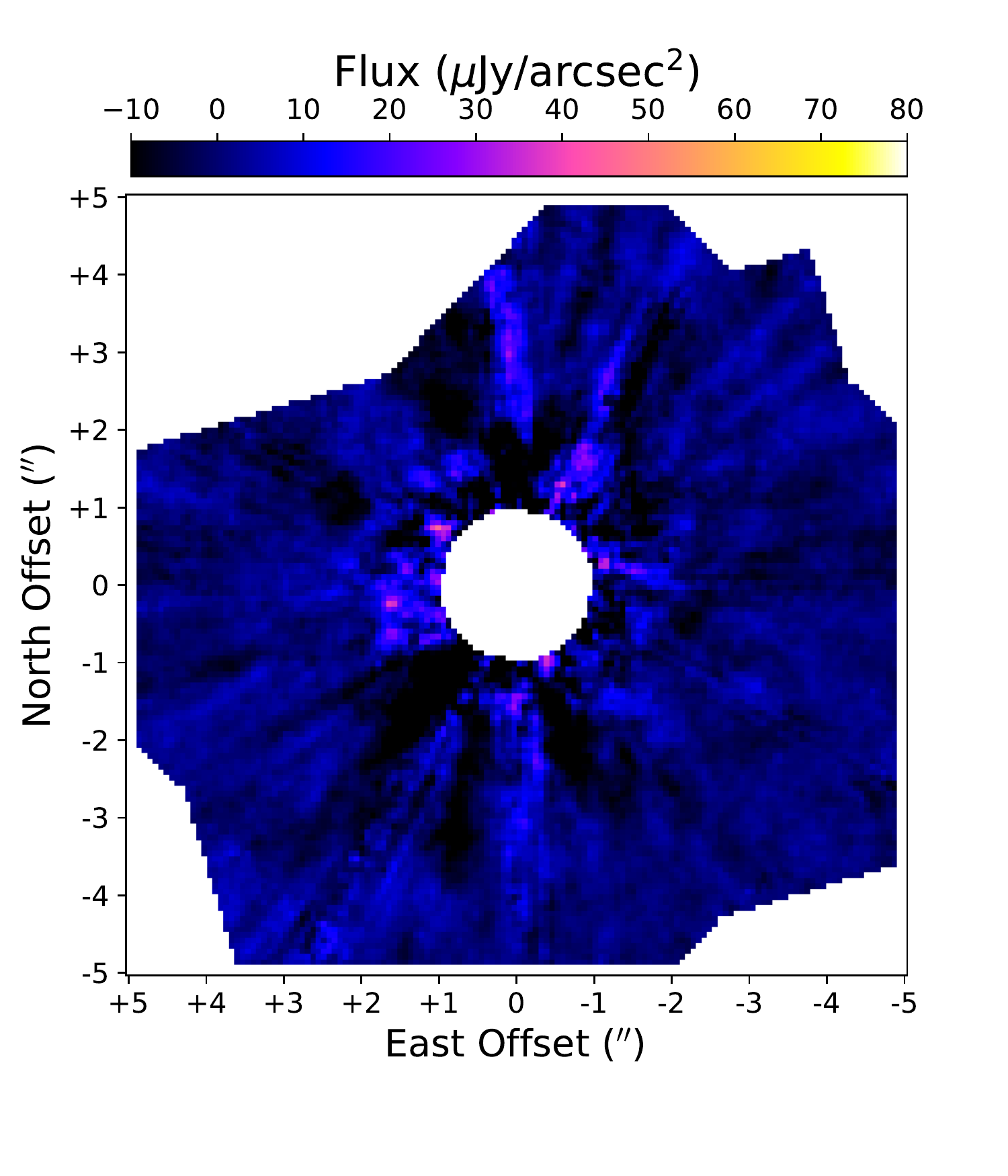}
	\includegraphics[width=0.33\textwidth]{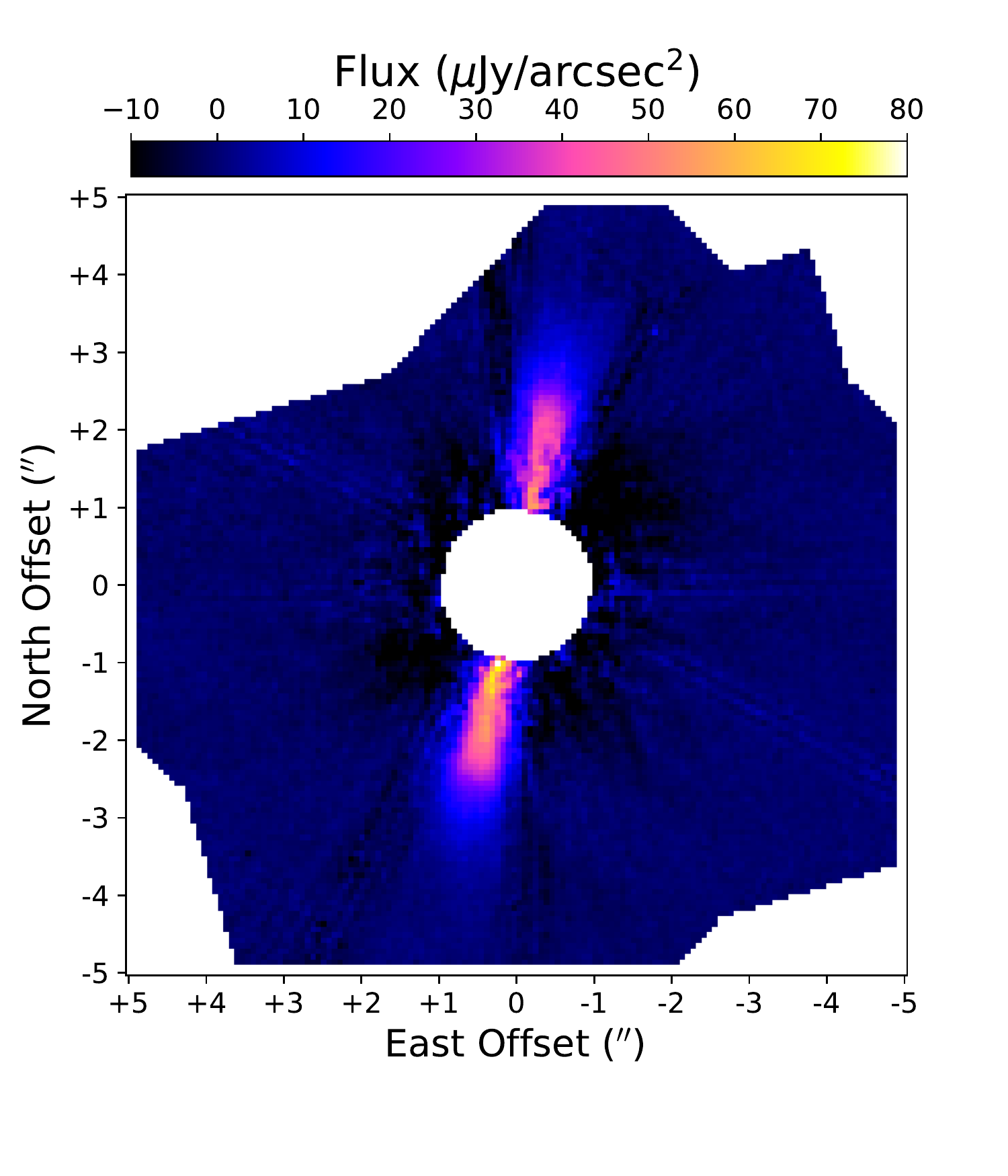}
	\vspace{-0.25in}
    \caption{\textit{HST}/NICMOS F110W (top) and F160W (bottom) observations of HD~16743. \textit{Left}: Reduced image. The centre of the image is dominated by residuals from the starlight suppression and has been masked in this plot to highlight fainter structure relevant to tracing the disc. Orientation is north up, east left. \textit{Middle}: Residual image showing the remnant noise after subtraction of the best-fit model disc from the observation. Artefacts are clearly visible at various position angles radiating from the central masked region. \textit{Right}: Scattered light model of the HD~16743 debris disc following the architecture determined from ALMA.}
    \label{fig:hd16743_hst-nicmos}
\end{figure*}

\subsubsection{VLT/SPHERE}

The star-subtracted HD~16743 obtained with VLT/SPHERE is shown in Figure \ref{fig:hd16743_vlt-sphere}, after applying an ADI reduction, smoothing with a square kernel of size 11 pixels (3.4 resolution elements) and masking the inner 1\farcs7, which is dominated by stellar residuals. There is a low S/N scattered light detection of the disc consistent with the ALMA image. The algorithm which turned out to best reveal this faint emission is a flavour of ADI known as radial ADI \cite[referred to as rADI in][]{Milli2012}, where the reference frame is estimated radially in annuli of width 1 resolution element and with a separation criteria $N_\delta=4$, meaning that frames with a field rotation smaller than 4 resolution elements at the separation of the considered annulus are not taken into account to limit self-subtraction effects \citep{Milli2012}. The image shown in Figure \ref{fig:hd16743_vlt-sphere} is expressed in $\mu\text{Jy}/\text{arcsec}^2$ but was not corrected by the throughput of the algorithm, meaning that the $\sim40~\mu\text{Jy}/\text{arcsec}^2$ of the faint emission compatible with the ALMA-resolved disc extent is a lower limit for the disc brightness.

\subsubsection{\textit{HST}/NICMOS}
To accurately estimate the albedo of the disc in the NICMOS images, we need to recover the disc photometry unbiased from the starlight subtraction artefacts produced by the PCA algorithm. To do so, we use a forward modelling approach that recover the fractions of the disc that are over- or self-subtracted by injecting negative disc models in the raw data and PCA-process them again. The residuals between the input model and the residual map reveals how the model was affected by the PSF subtraction process (forward model). 

We model the disc with an analytical optically thin disc model with height parameters in total, as used in \citet{Millar-Blanchaer2015}. The  radial dust density profile is modelled with two power laws set to indices $\beta_{in}=5$ and $\beta_{out}=-4$ inward and outward from a parent radius $R_{\rm break}$. The vertical dust density profile follows a Gaussian distribution with a standard deviation increasing linearly with the radius with an aspect ratio set to $h=0.05$, as expected for collisionally unperturbed debris discs \citep{Thebault2009}. The disc scattered-light brightness is modelled using a Henyey-Greenstein scattering phase function of parameter $g$, and the projected image of the disc is produced with an inclination $i$ and position angle $\phi$ set to $168\fdg6$ from the ALMA data best model. The disc brightness is then scaled with an arbitrary flux factor $f_{\rm disc}$.

To find the model that best fits both NICMOS images in F110W and F160W, we adjust five of these parameters: the peak radius $R_{\rm break}$, the vertical scale height $h$, the inclination $i$, the scattering phase function parameter $g$, and the scaling factor $f_{\rm disc}$. We use {\sc emcee} to identify the maximum likelihood parameters and their associated uncertainties. We used uniform priors for all five parameters. The {\sc emcee} process is initiated with 10 walkers per free parameter (50 total), and allowed to evolve for 7\,000 steps (350\,000 total realisations of the model). The walkers were initialised to values randomly selected around the ALMA best fit values for $R_{\rm break}$ and $i$, and around $h=0.05$ and $g=0.1$. Visual inspection of the chains and auto-correlation times suggests the runs are well converged after 200 steps. We draw the posterior probability distributions from every 16 steps of the converged chains. For the F110W image, the maximum probability model parameter values are $R_{\rm break} = 181.1~\pm~1.2~$au (assuming a distance of 57.806~pc), $h = 0.099~\pm~0.001$, $g = 0.704_{-0.006}^{+0.006}$, and $i = 89\fdg987_{-0\fdg005}^{+0\fdg002}$. For the F160W image, the best parameters are $R_{\rm break} = 153.0_{-2.2}^{+1.8}~$au, $g = 0.552_{-0.018}^{+0.015}$, $h = 0.112_{-0.003}^{+0.002}$, and $i = 89\fdg976_{-0\fdg028}^{+0\fdg011}$.

In Figure \ref{fig:hd16743_hst-nicmos}, we present the \textit{HST}/NICMOS F110W observations of HD~16743 along with the maximum probability model and the associated residuals. This is the higher snr of the two filter bands, and the F160W image shows similar structure. From the \textit{HST} data we find that the peak in scattered light and continuum emission for the disc are consistent within uncertainties. We find the maximum probability model suggests moderate forward scattering from the disc, consistent with other edge-on systems \citep[e.g.][]{2018Choquet}.

Following the method laid out in the previous sub-section, we calculate the albedo of the dust grains responsible for the scattered light disc from the posterior distribution of the fitting process. 
For the stellar flux, we adopt values of 6.1225~Jy in the F110W filter and 4.3092~Jy in the F160W filter, computed from the {\sc PySynphot} package \citep{Pysynphot2013} using a 7200~K and $\log(g)=4.4$ Kurucz spectrum normalised to HD~16743 $H$ band magnitude $H~=~5.971$. We derive albedo values of $\omega = 0.203~\pm~0.002$ in F110W, and $\omega = 0.286~\pm~0.006$ in F160W.

\section{Discussion}
\label{sec:discussion}

The high-resolution ALMA imaging presented here exhibits no structures indicative of stirring by a planetary companion; we find no evidence of a star-disc positional offset, nor any deviation of the disc from an axisymmetric ring. The large spatially resolved extent of HD~16743's disc, combined with its young age, was previously considered suggestive of stirring by a planetary companion \citep{2015Moor}. The then-current stirring models considered by \cite{2015Moor} could not account for such a system through self-stirring alone, subject to assumptions on the required planetesimal size \citep{2008KenyonBromley}. More recent self-stirring models require a smaller initial planetesimal size to trigger the collisional cascade \citep{2018KrivovBooth}. In this revised scenario, the state of HD~16743's disc can be explained through self-stirring without invoking an external perturber. The disc is also radially broad (FWHM $\simeq$ 79.4~au at 157.7~au, or $\Delta R/R \simeq 0.5$); re-observation at higher angular resolution would measure the disc radial profile to search for sculpting, or substructure to the belt at scales beneath the resolution reached with the present observations ($\leq$ 36~au). This would provide a further avenue to constrain the presence of planetary companions close to, or within the debris disc, but such observations would be hampered by the disc's edge-on orientation.

We have also imaged the disc in scattered light, obtaining a marginal detection with VLT/SPHERE $H$-band observations, and imaging the ansae of the disc in reprocessed archival \textit{HST}/NICMOS F110W and F160W observations. The dust albedoes derived from the scattered light brightnesses are consistent, pointing toward a scattering albedo of around 0.2 for the disc, consistent with similar systems \citep{2018Choquet,2018Marshall}. Such a low albedo is inconsistent with the expectations of Mie theory and astronomical silicate composition for the dust grains. The disc exhibits moderate forward scattering from the dust which is consistent with the relatively large minimum grain size of 5~$\mu$m inferred from the continuum emission. The disc shows the same extended vertical structure in scattered light as seen at millimetre wavelengths. We might expect the scattered light disc to be more extended due to the impact of stellar wind and radiation forces on the distribution of smaller dust grains. In this instance higher angular resolution observations at millimetre wavelengths to better resolve the vertical distribution of large grains (and the underlying planetesimals) within the system would clearly ascertain if this similarity is coincidence or not.

The millimetre and scattered light modelling of the disc reveals a preference for the vertical extent of the disc, measured to be 0.131$^{+0.014}_{-0.016}$, to be more extended than the typically assumed value of 0.05. We can use the scale height as a constraint on the mass of bodies perturbing the disc, following the analyses presented in \citep{2019Daley} and \citep{2019aMatra}. Taking the stellar mass of 1.537~$M_{\odot}$, a radius of 157.7~au for the disc, and the aforementioned scale height, we can apply these values to obtain a relative velocity of 1.6~$\pm$~0.2~km/s. Assuming the relative velocity is comparable to the escape velocity of the largest bodies in the planetesimal belt \citep{2014Schlichting}, and assuming a density of 2~g/cm$^{3}$ for those bodies consistent with massive trans-Neptunian objects \citep{2012Carry}, we can infer a mass of $0.005\pm0.001~{\rm M}_{\oplus}$ and corresponding radius of $1500^{+160}_{-180}$~km. This lower limit to the mass of perturbing bodies in the belt is around twice the mass of Pluto. 

Furthermore, we can calculate an upper limit to the mass of a perturbing body using the steady state model of \cite{2012Pan} using their equation 16 (and the formulation for $v_{\rm esc} > v_{r} > v_{R}$), following the reasoning of \citet{2019Daley}. They assumed an equilibrium between the stirring by large bodies and damping through collisions in the AU Mic disc, equating the velocity distribution of dust grains inferred from the scale height at millimetre wavelengths to the velocity distribution of the underlying planetesimal population stirring the disc and causing dust-producing collisions. Here, for the case of HD~16743, we assume that the stellar mass is $1.537~M_{\odot}$, the dust mass (in grains up to 1~mm) is $4.38\times10^{-3}~M_{\oplus}$. The velocity distribution of the dust grains is taken from the disc scale height, where $v_{\rm dust} = h \times v_{\rm Kepler}$. We thus obtain a joint limit on the number and mass of bodies perturbing the disc of $\sqrt{N}M = 18.3~M_{\oplus}$. In the limit of $N = 1$, we can therefore conclude that HD~16743's disc is being perturbed by bodies at least several times more massive than Pluto, and the maximum mass of a single body perturbing the belt is comparable to that of Neptune \textbf{($\simeq30~M_{\oplus}$)}.

\begin{figure}
    \centering
    \includegraphics[width=0.48\textwidth]{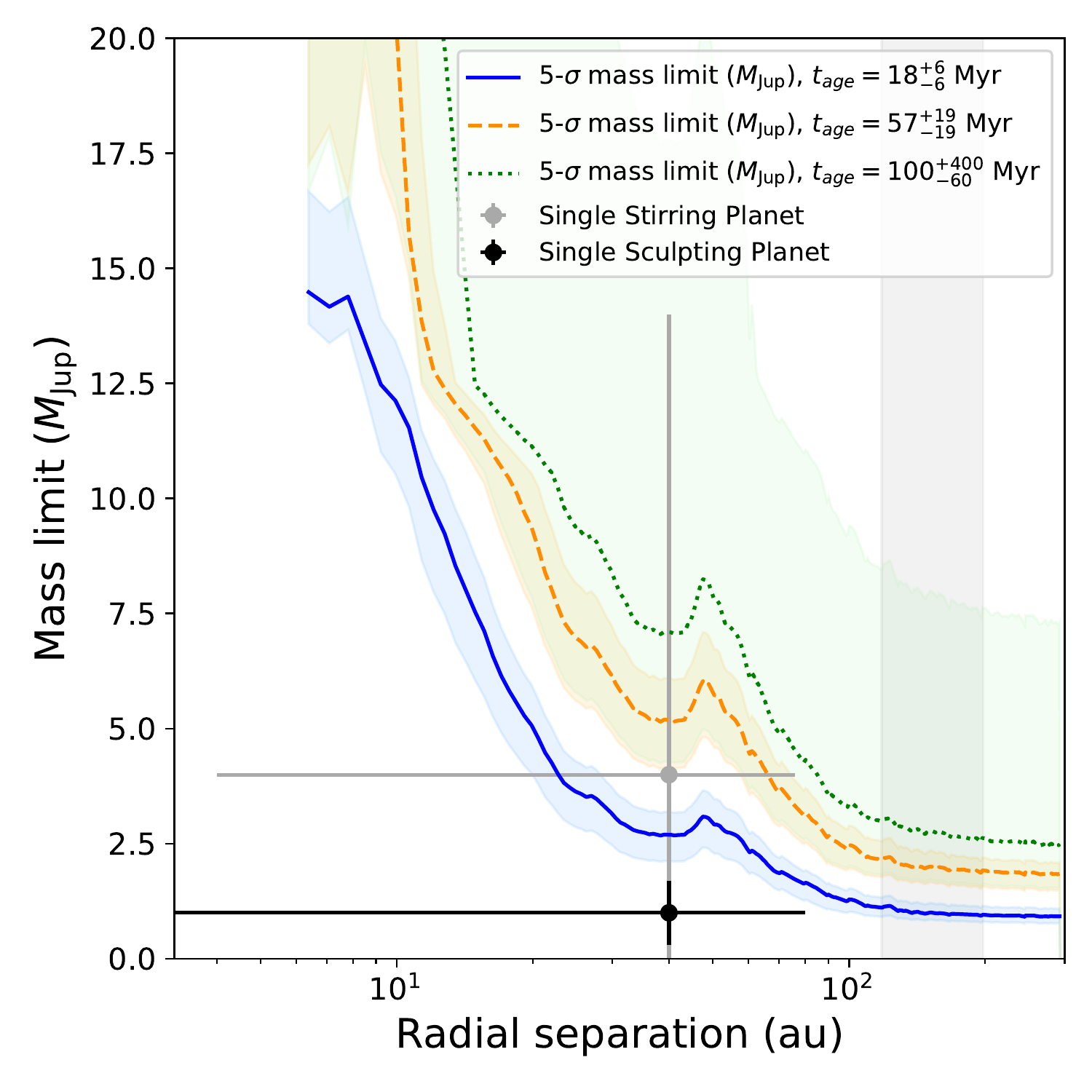}
    \caption{Upper limits to the mass of planetary companions to HD~16743 as a function of stellar separation. We convert the sensitivity from flux to mass for system ages (in Ma) of $18\pm6$ (blue, solid), $57\pm19$ (orange, dashed), and $100^{+400}_{-60}$ (green, dotted) using the evolutionary models of \citet{Baraffe2003}. The upper and lower uncertainties denoted by the shaded regions correspond to the uncertainties on the stellar ages. The grey shaded region denotes the approximate location of the disc planetesimal belt ($R_{\rm peak} \pm R_{\rm HWHM}$). The data points denote the companion masses required to sculpt the inner edge of the belt (black, $M_{\rm plt} = 1.1~\pm~0.7~M_{\rm Jup}$), or stir the disc (grey, $M_{\rm plt} = 4~\pm~10~M_{\rm Jup}$), respectively \citep{2022Pearce}.}
    \label{fig:mass_limit}
\end{figure}

We can also derive a mass limit to any sub-stellar companions from the point source sensitivity of the SPHERE map. We computed the detection limits using the  Andromeda algorithm \citep{Cantalloube2015} as implemented in the SPHERE Data Center \citep{Delorme2017,Galicher2018}. The conversion from flux density to mass was done using the AMES-COND evolutionary models \citep{Baraffe2003} assuming ages of $18\pm6$, $57\pm19$, and $100^{+400}_{-60}$~Ma consistent with the modelling presented in Section \ref{sec:star_params}. The resulting sensitivity map is relatively featureless, so we present the mass limits as a function of angular separation from the star in Figure \ref{fig:mass_limit}. The mass limits obtained here provide some additional constraint on the mass of any companion stirring the disc in the single planet limit as calculated by \citet{2022Pearce}, in the cases where the assumed age of the system is below 100~Ma. The multi-planet scenario from that work is consistent with the upper limit to a single mass stirring the disc derived in the previous paragraph.

\section{Conclusions}
\label{sec:conclusions}

We have spatially resolved the large, bright debris disc around HD~16743 in scattered light and millimetre wavelength imaging for the first time. The ALMA Band 6 interferometric observations reveal the architecture of the disc's outer belt to be well represented by a single, axisymmetric component with a radial extent, $R_{\rm peak} = 157.7^{+2.7}_{-1.5}~$au, consistent with previous far-infrared imaging observations, and a relatively broad width, $R_{\rm fwhm} / R_{\rm peak} = 0.5$. The disc lies in a near edge-on orientation with $i = 87\fdg3^{+1\fdg9}_{-2\fdg5}$, which is much more steeply inclined than previous estimates. Emission associated with an inner belt, inferred from the shape of the SED, is not detected in the ALMA observations, but this is consistent with the predicted emission based on standard assumptions. We also find no evidence of molecular CO gas emission from the disc, which is likewise consistent with the predicted non-detection based on models \citep{2017Kral}.

We calculate the sub-millimetre slope of the dust emission for HD~16743's disc to have an exponent of 3.72$~\pm~$0.03, significantly steeper than the oft-assumed steady state collisional cascade of 3.5 \citep{1969Dohnanyi}, and the bulk of debris disc measurements $\simeq~3.3$ \citep{2016Macgregor,2017Marshall,2021Norfolk}. Comparison with collisional models suggests the measured slope is consistent with the dust originating from collisions between bodies held together by their rigid strength \citep{2012Pan,2012Gaspar}.

The disc was observed in scattered light by both VLT/SPHERE and \textit{HST}/NICMOS. Coincidentally, both sets of observations had the same integration time (40~mins). The difference in quality between the detection with VLT/SPHERE and that obtained by \textit{HST}/NICMOS clearly demonstrates the power of space-based high contrast imaging to detect and image faint circumstellar discs, for systems where neither a high spatial resolution nor a small inner working angle are required.

We measured the disc albedo at near-infrared wavelengths with \textit{HST}/NICMOS, finding values of $\omega = 0.20$ in F110W and 0.29 in F160W. These values are consistent with observations of other debris disc systems \citep[e.g.][]{2018Choquet}, but substantially lower than theoretical expectations. We find no evidence of strong forward scattering from the dust grains, which is somewhat unusual, but this may be attributed to the disc centre being occulted by the coronagraphic mask in the \textit{HST}/NICMOS images, leaving the scattering properties to be derived only from the disc ansae.

The disc vertical scale height was resolved by both ALMA and \textit{HST}/NICMOS, with a value of 0.131$^{+0.014}_{-0.016}$ from the modelling the ALMA data. We use a simple analysis to relate the vertical extent of the disc to the mass of the underlying body or bodies perturbing the dust, obtaining limits in the range 0.005 to 18.3~$M_{\oplus}$. These constraints are much tighter than previously obtained limits of around a Jupiter mass for the companion to the disc based on \textit{Herschel} far-infrared imaging observations \citep{2022Pearce}.

No evidence for disc-planet interaction is found in the disc radial architecture at millimetre wavelengths, consistent with the expectation that the disc is self-stirred, following recent dynamical models \citep{2018KrivovBooth}. However, the broad fractional width of the disc provides an avenue for further exploration of this idea, as high resolution imaging of the majority of broad debris belts have revealed substructures consistent with gaps carved by a planetary companion \citep[e.g.][]{2018Marino,2019Marino,2019Macgregor}. Additionally, the edge-on orientation of the disc makes more detailed examination of the disc vertical scale height very feasible, providing another pathway to search for a low-mass companion stirring the debris disc.

\section*{Acknowledgements}

The authors thank the referee for their detailed and helpful comments which greatly improved the manuscript.

This research has made use of the SIMBAD database, operated at CDS, Strasbourg, France. This research has also made use of NASA's Astrophysics Data System.

JPM acknowledges research support by the Ministry of Science and Technology of Taiwan under grants MOST107-2119-M-001-031-MY3 and MOST109-2112-M-001-036-MY3, and Academia Sinica under grant AS-IA-106-M03.

CdB acknowledges support by Mexican CONACYT research grant FOP16-2021-01-320608.

This work was also partly supported by the Spanish program Unidad de Excelencia Mar\'ia de Maeztu CEX2020-001058-M, financed by MCIN/AEI/10.13039/501100011033.

GMK is supported by the Royal Society as a Royal Society University Research Fellow.

This paper makes use of the following ALMA data: ADS/JAO.ALMA\#2019.1.01220.S. ALMA is a partnership of ESO (representing its member states), NSF (USA) and NINS (Japan), together with NRC (Canada), MOST and ASIAA (Taiwan), and KASI (Republic of Korea), in cooperation with the Republic of Chile.

This work has made use of the SPHERE Data Centre, jointly operated by OSUG/IPAG (Grenoble), PYTHEAS/LAM/CeSAM (Marseille), OCA/Lagrange (Nice), Observatoire de Paris/LESIA (Paris), and Observatoire de Lyon/CRAL, and supported by a grant from Labex OSUG@2020 (Investissements d’avenir – ANR10 LABX56).

\textit{Software:} This paper has made use of the Python packages {\sc Astropy} \citep{2013AstroPy,2018AstroPy}, {\sc Corner} \citep{2016ForemanMackey}, {\sc Emcee} \citep{2013ForemanMackey}, {\sc Matplotlib} \citep{2007Hunter}, {\sc MiePython}, {\sc NumPy} \citep{2020Harris}, and {\sc SciPy} \citep{2020Virtanen}.

\textit{Facilities:} ALMA, \textit{HST}, VLT 

%%%%%%%%%%%%%%%%%%%%%%%%%%%%%%%%%%%%%%%%%%%%%%%%%%
\section*{Data Availability}

The data underlying this article are available in the article and in its online supplementary material.



\bibliographystyle{mnras}
\bibliography{hd16743.bbl} % if your bibtex file is called example.bib

\begin{thebibliography}{}
\makeatletter
\relax
\def\mn@urlcharsother{\let\do\@makeother \do\$\do\&\do\#\do\^\do\_\do\%\do\~}
\def\mn@doi{\begingroup\mn@urlcharsother \@ifnextchar [ {\mn@doi@}
  {\mn@doi@[]}}
\def\mn@doi@[#1]#2{\def\@tempa{#1}\ifx\@tempa\@empty \href
  {http://dx.doi.org/#2} {doi:#2}\else \href {http://dx.doi.org/#2} {#1}\fi
  \endgroup}
\def\mn@eprint#1#2{\mn@eprint@#1:#2::\@nil}
\def\mn@eprint@arXiv#1{\href {http://arxiv.org/abs/#1} {{\tt arXiv:#1}}}
\def\mn@eprint@dblp#1{\href {http://dblp.uni-trier.de/rec/bibtex/#1.xml}
  {dblp:#1}}
\def\mn@eprint@#1:#2:#3:#4\@nil{\def\@tempa {#1}\def\@tempb {#2}\def\@tempc
  {#3}\ifx \@tempc \@empty \let \@tempc \@tempb \let \@tempb \@tempa \fi \ifx
  \@tempb \@empty \def\@tempb {arXiv}\fi \@ifundefined
  {mn@eprint@\@tempb}{\@tempb:\@tempc}{\expandafter \expandafter \csname
  mn@eprint@\@tempb\endcsname \expandafter{\@tempc}}}

\bibitem[\protect\citeauthoryear{{Allard}, {Homeier}  \& {Freytag}}{{Allard}
  et~al.}{2011}]{2011Allard}
{Allard} F.,  {Homeier} D.,   {Freytag} B.,  2011, in {Johns-Krull} C.,
  {Browning} M.~K.,   {West} A.~A.,  eds,  Astronomical Society of the Pacific
  Conference Series Vol. 448, 16th Cambridge Workshop on Cool Stars, Stellar
  Systems, and the Sun. p.~91 (\mn@eprint {arXiv} {1011.5405})

\bibitem[\protect\citeauthoryear{{Allard}, {Homeier}  \& {Freytag}}{{Allard}
  et~al.}{2012}]{2012Allard}
{Allard} F.,  {Homeier} D.,   {Freytag} B.,  2012, \mn@doi [Philosophical
  Transactions of the Royal Society of London Series A]
  {10.1098/rsta.2011.0269}, \href
  {https://ui.adsabs.harvard.edu/abs/2012RSPTA.370.2765A} {370, 2765}

\bibitem[\protect\citeauthoryear{{Andrews} et~al.,}{{Andrews}
  et~al.}{2018}]{2018Andrews}
{Andrews} S.~M.,  et~al., 2018, \mn@doi [\apjl] {10.3847/2041-8213/aaf741},
  \href {https://ui.adsabs.harvard.edu/abs/2018ApJ...869L..41A} {869, L41}

\bibitem[\protect\citeauthoryear{{Asplund}, {Grevesse}, {Sauval}  \&
  {Scott}}{{Asplund} et~al.}{2009}]{2009Asplund}
{Asplund} M.,  {Grevesse} N.,  {Sauval} A.~J.,   {Scott} P.,  2009, \mn@doi
  [\araa] {10.1146/annurev.astro.46.060407.145222}, \href
  {https://ui.adsabs.harvard.edu/abs/2009ARA&A..47..481A} {47, 481}

\bibitem[\protect\citeauthoryear{{Astropy Collaboration} et~al.,}{{Astropy
  Collaboration} et~al.}{2013}]{2013AstroPy}
{Astropy Collaboration} et~al., 2013, \mn@doi [\aap]
  {10.1051/0004-6361/201322068}, \href
  {https://ui.adsabs.harvard.edu/#abs/2013A&A...558A..33A} {558, A33}

\bibitem[\protect\citeauthoryear{{Astropy Collaboration} et~al.,}{{Astropy
  Collaboration} et~al.}{2018}]{2018AstroPy}
{Astropy Collaboration} et~al., 2018, \mn@doi [\aj] {10.3847/1538-3881/aabc4f},
  \href {http://adsabs.harvard.edu/abs/2018AJ....156..123A} {156, 123}

\bibitem[\protect\citeauthoryear{{Baraffe}, {Chabrier}, {Barman}, {Allard}  \&
  {Hauschildt}}{{Baraffe} et~al.}{2003}]{Baraffe2003}
{Baraffe} I.,  {Chabrier} G.,  {Barman} T.~S.,  {Allard} F.,   {Hauschildt}
  P.~H.,  2003, \mn@doi [\aap] {10.1051/0004-6361:20030252}, \href
  {https://ui.adsabs.harvard.edu/abs/2003A&A...402..701B} {402, 701}

\bibitem[\protect\citeauthoryear{{Barber}, {Tennyson}, {Harris}  \&
  {Tolchenov}}{{Barber} et~al.}{2006}]{2006Barber}
{Barber} R.~J.,  {Tennyson} J.,  {Harris} G.~J.,   {Tolchenov} R.~N.,  2006,
  \mn@doi [\mnras] {10.1111/j.1365-2966.2006.10184.x}, \href
  {https://ui.adsabs.harvard.edu/abs/2006MNRAS.368.1087B} {368, 1087}

\bibitem[\protect\citeauthoryear{{Beuzit} et~al.,}{{Beuzit}
  et~al.}{2019}]{Beuzit2019}
{Beuzit} J.~L.,  et~al., 2019, \mn@doi [\aap] {10.1051/0004-6361/201935251},
  \href {https://ui.adsabs.harvard.edu/abs/2019A&A...631A.155B} {631, A155}

\bibitem[\protect\citeauthoryear{{Bressan}, {Marigo}, {Girardi}, {Salasnich},
  {Dal Cero}, {Rubele}  \& {Nanni}}{{Bressan} et~al.}{2012}]{2012Bressan}
{Bressan} A.,  {Marigo} P.,  {Girardi} L.,  {Salasnich} B.,  {Dal Cero} C.,
  {Rubele} S.,   {Nanni} A.,  2012, \mn@doi [\mnras]
  {10.1111/j.1365-2966.2012.21948.x}, \href
  {https://ui.adsabs.harvard.edu/abs/2012MNRAS.427..127B} {427, 127}

\bibitem[\protect\citeauthoryear{{Cantalloube} et~al.,}{{Cantalloube}
  et~al.}{2015}]{Cantalloube2015}
{Cantalloube} F.,  et~al., 2015, \mn@doi [\aap] {10.1051/0004-6361/201425571},
  \href {https://ui.adsabs.harvard.edu/abs/2015A&A...582A..89C} {582, A89}

\bibitem[\protect\citeauthoryear{{Cantat-Gaudin}}{{Cantat-Gaudin}}{2022}]{2022Cantat-Gaudin}
{Cantat-Gaudin} T.,  2022, \mn@doi [Universe] {10.3390/universe8020111}, \href
  {https://ui.adsabs.harvard.edu/abs/2022Univ....8..111C} {8, 111}

\bibitem[\protect\citeauthoryear{{Carry}}{{Carry}}{2012}]{2012Carry}
{Carry} B.,  2012, \mn@doi [\planss] {10.1016/j.pss.2012.03.009}, \href
  {https://ui.adsabs.harvard.edu/abs/2012P&SS...73...98C} {73, 98}

\bibitem[\protect\citeauthoryear{{Chambers}}{{Chambers}}{2001}]{2001Chambers}
{Chambers} J.~E.,  2001, \mn@doi [\icarus] {10.1006/icar.2001.6639}, \href
  {https://ui.adsabs.harvard.edu/abs/2001Icar..152..205C} {152, 205}

\bibitem[\protect\citeauthoryear{{Chambers}}{{Chambers}}{2004}]{2004Chambers}
{Chambers} J.~E.,  2004, \mn@doi [Earth and Planetary Science Letters]
  {10.1016/j.epsl.2004.04.031}, \href
  {https://ui.adsabs.harvard.edu/abs/2004E&PSL.223..241C} {223, 241}

\bibitem[\protect\citeauthoryear{{Choquet} et~al.,}{{Choquet}
  et~al.}{2014}]{Choquet2014}
{Choquet} {\'E}.,  et~al., 2014, in \procspie. p.~57 (\mn@eprint {arXiv}
  {1407.0617}), \mn@doi{10.1117/12.2056672}, \url
  {http://adsabs.harvard.edu/abs/2014SPIE.9143E..57C}

\bibitem[\protect\citeauthoryear{{Choquet} et~al.,}{{Choquet}
  et~al.}{2016}]{Choquet2016}
{Choquet} {\'E}.,  et~al., 2016, \mn@doi [\apjl] {10.3847/2041-8205/817/1/L2},
  817, L2

\bibitem[\protect\citeauthoryear{{Choquet} et~al.,}{{Choquet}
  et~al.}{2017}]{Choquet2017}
{Choquet} {\'E}.,  et~al., 2017, \mn@doi [\apjl] {10.3847/2041-8213/834/2/L12},
  \href {http://adsabs.harvard.edu/abs/2017ApJ...834L..12C} {834, L12}

\bibitem[\protect\citeauthoryear{{Choquet} et~al.,}{{Choquet}
  et~al.}{2018}]{2018Choquet}
{Choquet} {\'E}.,  et~al., 2018, \mn@doi [\apj] {10.3847/1538-4357/aaa892},
  \href {https://ui.adsabs.harvard.edu/abs/2018ApJ...854...53C} {854, 53}

\bibitem[\protect\citeauthoryear{{Cronin-Coltsmann} et~al.,}{{Cronin-Coltsmann}
  et~al.}{2021}]{Cronin-Coltsmann2021}
{Cronin-Coltsmann} P.~F.,  et~al., 2021, \mn@doi [\mnras]
  {10.1093/mnras/stab1237}, \href
  {https://ui.adsabs.harvard.edu/abs/2021MNRAS.504.4497C} {504, 4497}

\bibitem[\protect\citeauthoryear{{Daley} et~al.,}{{Daley}
  et~al.}{2019}]{2019Daley}
{Daley} C.,  et~al., 2019, \mn@doi [\apj] {10.3847/1538-4357/ab1074}, \href
  {https://ui.adsabs.harvard.edu/abs/2019ApJ...875...87D} {875, 87}

\bibitem[\protect\citeauthoryear{{Delorme} et~al.,}{{Delorme}
  et~al.}{2017}]{Delorme2017}
{Delorme} P.,  et~al., 2017, in {Reyl{\'e}} C.,  {Di Matteo} P.,  {Herpin} F.,
  {Lagadec} E.,  {Lan{\c{c}}on} A.,  {Meliani} Z.,   {Royer} F.,  eds,
  SF2A-2017: Proceedings of the Annual meeting of the French Society of
  Astronomy and Astrophysics. p.~Di (\mn@eprint {arXiv} {1712.06948})

\bibitem[\protect\citeauthoryear{{Desidera} et~al.,}{{Desidera}
  et~al.}{2015}]{Desidera2015}
{Desidera} S.,  et~al., 2015, \mn@doi [\aap] {10.1051/0004-6361/201323168},
  \href {https://ui.adsabs.harvard.edu/abs/2015A&A...573A.126D} {573, A126}

\bibitem[\protect\citeauthoryear{{Dohlen} et~al.,}{{Dohlen}
  et~al.}{2008}]{Dohlen2008}
{Dohlen} K.,  et~al., 2008, in Society of Photo-Optical Instrumentation
  Engineers (SPIE) Conference Series. , \mn@doi{10.1117/12.789786}

\bibitem[\protect\citeauthoryear{{Dohnanyi}}{{Dohnanyi}}{1969}]{1969Dohnanyi}
{Dohnanyi} J.~S.,  1969, \mn@doi [\jgr] {10.1029/JB074i010p02531}, \href
  {https://ui.adsabs.harvard.edu/abs/1969JGR....74.2531D} {74, 2531}

\bibitem[\protect\citeauthoryear{{Dong}, {Li}, {Chiang}  \& {Li}}{{Dong}
  et~al.}{2018}]{2018Dong}
{Dong} R.,  {Li} S.,  {Chiang} E.,   {Li} H.,  2018, \mn@doi [\apj]
  {10.3847/1538-4357/aadadd}, \href
  {https://ui.adsabs.harvard.edu/abs/2018ApJ...866..110D} {866, 110}

\bibitem[\protect\citeauthoryear{{Draine}}{{Draine}}{2003}]{2003Draine}
{Draine} B.~T.,  2003, \mn@doi [\apj] {10.1086/379118}, \href
  {https://ui.adsabs.harvard.edu/abs/2003ApJ...598.1017D} {598, 1017}

\bibitem[\protect\citeauthoryear{{Dullemond}, {Juhasz}, {Pohl}, {Sereshti},
  {Shetty}, {Peters}, {Commercon}  \& {Flock}}{{Dullemond}
  et~al.}{2012}]{2012Dullemond}
{Dullemond} C.~P.,  {Juhasz} A.,  {Pohl} A.,  {Sereshti} F.,  {Shetty} R.,
  {Peters} T.,  {Commercon} B.,   {Flock} M.,  2012, {RADMC-3D: A multi-purpose
  radiative transfer tool}, Astrophysics Source Code Library, record
  ascl:1202.015 (\mn@eprint {ascl} {1202.015})

\bibitem[\protect\citeauthoryear{{Esposito} et~al.,}{{Esposito}
  et~al.}{2020}]{2020Esposito}
{Esposito} T.~M.,  et~al., 2020, \mn@doi [\aj] {10.3847/1538-3881/ab9199},
  \href {https://ui.adsabs.harvard.edu/abs/2020AJ....160...24E} {160, 24}

\bibitem[\protect\citeauthoryear{{Faramaz} et~al.,}{{Faramaz}
  et~al.}{2019}]{2019Faramaz}
{Faramaz} V.,  et~al., 2019, \mn@doi [\aj] {10.3847/1538-3881/ab3ec1}, \href
  {https://ui.adsabs.harvard.edu/abs/2019AJ....158..162F} {158, 162}

\bibitem[\protect\citeauthoryear{{Faramaz} et~al.,}{{Faramaz}
  et~al.}{2021}]{2021Faramaz}
{Faramaz} V.,  et~al., 2021, \mn@doi [\aj] {10.3847/1538-3881/abf4e0}, \href
  {https://ui.adsabs.harvard.edu/abs/2021AJ....161..271F} {161, 271}

\bibitem[\protect\citeauthoryear{Foreman-Mackey}{Foreman-Mackey}{2016}]{2016ForemanMackey}
Foreman-Mackey D.,  2016, \mn@doi [The Journal of Open Source Software]
  {10.21105/joss.00024}, 24

\bibitem[\protect\citeauthoryear{{Foreman-Mackey}, {Hogg}, {Lang}  \&
  {Goodman}}{{Foreman-Mackey} et~al.}{2013}]{2013ForemanMackey}
{Foreman-Mackey} D.,  {Hogg} D.~W.,  {Lang} D.,   {Goodman} J.,  2013, \mn@doi
  [\pasp] {10.1086/670067}, \href
  {https://ui.adsabs.harvard.edu/abs/2013PASP..125..306F} {125, 306}

\bibitem[\protect\citeauthoryear{{Gaia Collaboration} et~al.,}{{Gaia
  Collaboration} et~al.}{2016}]{2016Gaia}
{Gaia Collaboration} et~al., 2016, \mn@doi [\aap]
  {10.1051/0004-6361/201629272}, \href
  {https://ui.adsabs.harvard.edu/abs/2016A&A...595A...1G} {595, A1}

\bibitem[\protect\citeauthoryear{{Gaia Collaboration} et~al.,}{{Gaia
  Collaboration} et~al.}{2018}]{2018Gaia}
{Gaia Collaboration} et~al., 2018, \mn@doi [\aap]
  {10.1051/0004-6361/201833051}, \href
  {https://ui.adsabs.harvard.edu/abs/2018A&A...616A...1G} {616, A1}

\bibitem[\protect\citeauthoryear{{Gaia Collaboration} et~al.,}{{Gaia
  Collaboration} et~al.}{2021}]{2021Gaia}
{Gaia Collaboration} et~al., 2021, \mn@doi [\aap]
  {10.1051/0004-6361/202039657}, \href
  {https://ui.adsabs.harvard.edu/abs/2021A&A...649A...1G} {649, A1}

\bibitem[\protect\citeauthoryear{{Galicher} et~al.,}{{Galicher}
  et~al.}{2018}]{Galicher2018}
{Galicher} R.,  et~al., 2018, \mn@doi [\aap] {10.1051/0004-6361/201832973},
  \href {https://ui.adsabs.harvard.edu/abs/2018A&A...615A..92G} {615, A92}

\bibitem[\protect\citeauthoryear{{G{\'a}sp{\'a}r}, {Psaltis}, {Rieke}  \&
  {{\"O}zel}}{{G{\'a}sp{\'a}r} et~al.}{2012}]{2012Gaspar}
{G{\'a}sp{\'a}r} A.,  {Psaltis} D.,  {Rieke} G.~H.,   {{\"O}zel} F.,  2012,
  \mn@doi [\apj] {10.1088/0004-637X/754/1/74}, \href
  {https://ui.adsabs.harvard.edu/abs/2012ApJ...754...74G} {754, 74}

\bibitem[\protect\citeauthoryear{{Geiler}, {Krivov}, {Booth}  \&
  {L{\"o}hne}}{{Geiler} et~al.}{2019}]{2019Geiler}
{Geiler} F.,  {Krivov} A.~V.,  {Booth} M.,   {L{\"o}hne} T.,  2019, \mn@doi
  [\mnras] {10.1093/mnras/sty3160}, \href
  {https://ui.adsabs.harvard.edu/abs/2019MNRAS.483..332G} {483, 332}

\bibitem[\protect\citeauthoryear{{Greaves} et~al.,}{{Greaves}
  et~al.}{2016}]{2016Greaves}
{Greaves} J.~S.,  et~al., 2016, \mn@doi [\mnras] {10.1093/mnras/stw1569}, \href
  {https://ui.adsabs.harvard.edu/abs/2016MNRAS.461.3910G} {461, 3910}

\bibitem[\protect\citeauthoryear{{Hagan}, {Choquet}, {Soummer}  \&
  {Vigan}}{{Hagan} et~al.}{2018}]{Hagan2018}
{Hagan} J.~B.,  {Choquet} {\'E}.,  {Soummer} R.,   {Vigan} A.,  2018, \mn@doi
  [\aj] {10.3847/1538-3881/aab14b}, 155, 179

\bibitem[\protect\citeauthoryear{{Hales} et~al.,}{{Hales}
  et~al.}{2022}]{2022Hales}
{Hales} A.~S.,  et~al., 2022, \mn@doi [\apj] {10.3847/1538-4357/ac9cd3}, \href
  {https://ui.adsabs.harvard.edu/abs/2022ApJ...940..161H} {940, 161}

\bibitem[\protect\citeauthoryear{Harris et~al.,}{Harris
  et~al.}{2020}]{2020Harris}
Harris C.~R.,  et~al., 2020, \mn@doi [Nature] {10.1038/s41586-020-2649-2}, 585,
  357

\bibitem[\protect\citeauthoryear{{H{\o}g} et~al.,}{{H{\o}g}
  et~al.}{2000}]{2000Hog}
{H{\o}g} E.,  et~al., 2000, \aap, \href
  {https://ui.adsabs.harvard.edu/abs/2000A&A...355L..27H} {355, L27}

\bibitem[\protect\citeauthoryear{{Huang} et~al.,}{{Huang}
  et~al.}{2018}]{2018Huang}
{Huang} J.,  et~al., 2018, \mn@doi [\apjl] {10.3847/2041-8213/aaf740}, \href
  {https://ui.adsabs.harvard.edu/abs/2018ApJ...869L..42H} {869, L42}

\bibitem[\protect\citeauthoryear{{Hughes}, {Duch{\^e}ne}  \&
  {Matthews}}{{Hughes} et~al.}{2018}]{2018Hughes}
{Hughes} A.~M.,  {Duch{\^e}ne} G.,   {Matthews} B.~C.,  2018, \mn@doi [\araa]
  {10.1146/annurev-astro-081817-052035}, \href
  {https://ui.adsabs.harvard.edu/abs/2018ARA&A..56..541H} {56, 541}

\bibitem[\protect\citeauthoryear{{Hunter}}{{Hunter}}{2007}]{2007Hunter}
{Hunter} J.~D.,  2007, \mn@doi [Computing in Science and Engineering]
  {10.1109/MCSE.2007.55}, \href
  {https://ui.adsabs.harvard.edu/#abs/2007CSE.....9...90H} {9, 90}

\bibitem[\protect\citeauthoryear{{Ishihara} et~al.,}{{Ishihara}
  et~al.}{2010}]{2010Ishihara}
{Ishihara} D.,  et~al., 2010, \mn@doi [\aap] {10.1051/0004-6361/200913811},
  \href {https://ui.adsabs.harvard.edu/abs/2010A&A...514A...1I} {514, A1}

\bibitem[\protect\citeauthoryear{{Johansen}, {Ronnet}, {Schiller}, {Deng}  \&
  {Bizzarro}}{{Johansen} et~al.}{2023}]{2023Johansen}
{Johansen} A.,  {Ronnet} T.,  {Schiller} M.,  {Deng} Z.,   {Bizzarro} M.,
  2023, \mn@doi [\aap] {10.1051/0004-6361/202142141}, \href
  {https://ui.adsabs.harvard.edu/abs/2023A&A...671A..74J} {671, A74}

\bibitem[\protect\citeauthoryear{{Kennedy}}{{Kennedy}}{2020}]{2020Kennedy}
{Kennedy} G.~M.,  2020, \mn@doi [Royal Society Open Science]
  {10.1098/rsos.200063}, \href
  {https://ui.adsabs.harvard.edu/abs/2020RSOS....700063K} {7, 200063}

\bibitem[\protect\citeauthoryear{{Kennedy}, {Wyatt}, {Sibthorpe}, {Phillips},
  {Matthews}  \& {Greaves}}{{Kennedy} et~al.}{2012}]{2012Kennedy}
{Kennedy} G.~M.,  {Wyatt} M.~C.,  {Sibthorpe} B.,  {Phillips} N.~M.,
  {Matthews} B.~C.,   {Greaves} J.~S.,  2012, \mn@doi [\mnras]
  {10.1111/j.1365-2966.2012.21865.x}, \href
  {https://ui.adsabs.harvard.edu/abs/2012MNRAS.426.2115K} {426, 2115}

\bibitem[\protect\citeauthoryear{{Kenyon} \& {Bromley}}{{Kenyon} \&
  {Bromley}}{2008}]{2008KenyonBromley}
{Kenyon} S.~J.,  {Bromley} B.~C.,  2008, \mn@doi [\apjs] {10.1086/591794},
  \href {https://ui.adsabs.harvard.edu/abs/2008ApJS..179..451K} {179, 451}

\bibitem[\protect\citeauthoryear{{Kokubo} \& {Ida}}{{Kokubo} \&
  {Ida}}{1998}]{1998Kokubo}
{Kokubo} E.,  {Ida} S.,  1998, \mn@doi [\icarus] {10.1006/icar.1997.5840},
  \href {https://ui.adsabs.harvard.edu/abs/1998Icar..131..171K} {131, 171}

\bibitem[\protect\citeauthoryear{{Kral}, {Matr{\`a}}, {Wyatt}  \&
  {Kennedy}}{{Kral} et~al.}{2017}]{2017Kral}
{Kral} Q.,  {Matr{\`a}} L.,  {Wyatt} M.~C.,   {Kennedy} G.~M.,  2017, \mn@doi
  [\mnras] {10.1093/mnras/stx730}, \href
  {https://ui.adsabs.harvard.edu/abs/2017MNRAS.469..521K} {469, 521}

\bibitem[\protect\citeauthoryear{{Kral}, {Matr{\`a}}, {Kennedy}, {Marino}  \&
  {Wyatt}}{{Kral} et~al.}{2020}]{2020Kral}
{Kral} Q.,  {Matr{\`a}} L.,  {Kennedy} G.~M.,  {Marino} S.,   {Wyatt} M.~C.,
  2020, \mn@doi [\mnras] {10.1093/mnras/staa2038}, \href
  {https://ui.adsabs.harvard.edu/abs/2020MNRAS.497.2811K} {497, 2811}

\bibitem[\protect\citeauthoryear{{Krivov} \& {Booth}}{{Krivov} \&
  {Booth}}{2018}]{2018KrivovBooth}
{Krivov} A.~V.,  {Booth} M.,  2018, \mn@doi [\mnras] {10.1093/mnras/sty1607},
  \href {https://ui.adsabs.harvard.edu/abs/2018MNRAS.479.3300K} {479, 3300}

\bibitem[\protect\citeauthoryear{{Krivov} \& {Wyatt}}{{Krivov} \&
  {Wyatt}}{2021}]{2021KrivovWyatt}
{Krivov} A.~V.,  {Wyatt} M.~C.,  2021, \mn@doi [\mnras]
  {10.1093/mnras/staa2385}, \href
  {https://ui.adsabs.harvard.edu/abs/2021MNRAS.500..718K} {500, 718}

\bibitem[\protect\citeauthoryear{{Lebouteiller}, {Barry}, {Spoon},
  {Bernard-Salas}, {Sloan}, {Houck}  \& {Weedman}}{{Lebouteiller}
  et~al.}{2011}]{2011Lebouteiller}
{Lebouteiller} V.,  {Barry} D.~J.,  {Spoon} H.~W.~W.,  {Bernard-Salas} J.,
  {Sloan} G.~C.,  {Houck} J.~R.,   {Weedman} D.~W.,  2011, \mn@doi [\apjs]
  {10.1088/0067-0049/196/1/8}, \href
  {https://ui.adsabs.harvard.edu/abs/2011ApJS..196....8L} {196, 8}

\bibitem[\protect\citeauthoryear{{Lodato} et~al.,}{{Lodato}
  et~al.}{2019}]{2019Lodato}
{Lodato} G.,  et~al., 2019, \mn@doi [\mnras] {10.1093/mnras/stz913}, \href
  {https://ui.adsabs.harvard.edu/abs/2019MNRAS.486..453L} {486, 453}

\bibitem[\protect\citeauthoryear{{Long} et~al.,}{{Long}
  et~al.}{2018}]{2018Long}
{Long} F.,  et~al., 2018, \mn@doi [\apj] {10.3847/1538-4357/aae8e1}, \href
  {https://ui.adsabs.harvard.edu/abs/2018ApJ...869...17L} {869, 17}

\bibitem[\protect\citeauthoryear{{Lovell} et~al.,}{{Lovell}
  et~al.}{2021}]{2021Lovell}
{Lovell} J.~B.,  et~al., 2021, \mn@doi [\mnras] {10.1093/mnras/staa3335}, \href
  {https://ui.adsabs.harvard.edu/abs/2021MNRAS.500.4878L} {500, 4878}

\bibitem[\protect\citeauthoryear{{MacGregor} et~al.,}{{MacGregor}
  et~al.}{2016}]{2016Macgregor}
{MacGregor} M.~A.,  et~al., 2016, \mn@doi [\apj] {10.3847/0004-637X/823/2/79},
  \href {https://ui.adsabs.harvard.edu/abs/2016ApJ...823...79M} {823, 79}

\bibitem[\protect\citeauthoryear{{MacGregor} et~al.,}{{MacGregor}
  et~al.}{2017}]{2017Macgregor}
{MacGregor} M.~A.,  et~al., 2017, \mn@doi [\apj] {10.3847/1538-4357/aa71ae},
  \href {https://ui.adsabs.harvard.edu/abs/2017ApJ...842....8M} {842, 8}

\bibitem[\protect\citeauthoryear{{MacGregor} et~al.,}{{MacGregor}
  et~al.}{2018}]{2018Macgregor}
{MacGregor} M.~A.,  et~al., 2018, \mn@doi [\apj] {10.3847/1538-4357/aaec71},
  \href {https://ui.adsabs.harvard.edu/abs/2018ApJ...869...75M} {869, 75}

\bibitem[\protect\citeauthoryear{{MacGregor} et~al.,}{{MacGregor}
  et~al.}{2019}]{2019Macgregor}
{MacGregor} M.~A.,  et~al., 2019, \mn@doi [\apjl] {10.3847/2041-8213/ab21c2},
  \href {https://ui.adsabs.harvard.edu/abs/2019ApJ...877L..32M} {877, L32}

\bibitem[\protect\citeauthoryear{{Manara}, {Morbidelli}  \& {Guillot}}{{Manara}
  et~al.}{2018}]{2018Manara}
{Manara} C.~F.,  {Morbidelli} A.,   {Guillot} T.,  2018, \mn@doi [\aap]
  {10.1051/0004-6361/201834076}, \href
  {https://ui.adsabs.harvard.edu/abs/2018A&A...618L...3M} {618, L3}

\bibitem[\protect\citeauthoryear{{Marino}}{{Marino}}{2021}]{2021Marino}
{Marino} S.,  2021, \mn@doi [\mnras] {10.1093/mnras/stab771}, \href
  {https://ui.adsabs.harvard.edu/abs/2021MNRAS.503.5100M} {503, 5100}

\bibitem[\protect\citeauthoryear{{Marino} et~al.,}{{Marino}
  et~al.}{2018}]{2018Marino}
{Marino} S.,  et~al., 2018, \mn@doi [\mnras] {10.1093/mnras/sty1790}, \href
  {https://ui.adsabs.harvard.edu/abs/2018MNRAS.479.5423M} {479, 5423}

\bibitem[\protect\citeauthoryear{{Marino}, {Yelverton}, {Booth}, {Faramaz},
  {Kennedy}, {Matr{\`a}}  \& {Wyatt}}{{Marino} et~al.}{2019}]{2019Marino}
{Marino} S.,  {Yelverton} B.,  {Booth} M.,  {Faramaz} V.,  {Kennedy} G.~M.,
  {Matr{\`a}} L.,   {Wyatt} M.~C.,  2019, \mn@doi [\mnras]
  {10.1093/mnras/stz049}, \href
  {https://ui.adsabs.harvard.edu/abs/2019MNRAS.484.1257M} {484, 1257}

\bibitem[\protect\citeauthoryear{{Marino}, {Flock}, {Henning}, {Kral},
  {Matr{\`a}}  \& {Wyatt}}{{Marino} et~al.}{2020}]{2020Marino}
{Marino} S.,  {Flock} M.,  {Henning} T.,  {Kral} Q.,  {Matr{\`a}} L.,   {Wyatt}
  M.~C.,  2020, \mn@doi [\mnras] {10.1093/mnras/stz3487}, \href
  {https://ui.adsabs.harvard.edu/abs/2020MNRAS.492.4409M} {492, 4409}

\bibitem[\protect\citeauthoryear{{Marois}, {Lafreni{\`e}re}, {Doyon},
  {Macintosh}  \& {Nadeau}}{{Marois} et~al.}{2006}]{Marois2006}
{Marois} C.,  {Lafreni{\`e}re} D.,  {Doyon} R.,  {Macintosh} B.,   {Nadeau} D.,
   2006, \mn@doi [\apj] {10.1086/500401}, \href
  {http://cdsads.u-strasbg.fr/abs/2006ApJ...641..556M} {641, 556}

\bibitem[\protect\citeauthoryear{{Marshall} et~al.,}{{Marshall}
  et~al.}{2014}]{2014Marshall}
{Marshall} J.~P.,  et~al., 2014, \mn@doi [\aap] {10.1051/0004-6361/201424517},
  \href {https://ui.adsabs.harvard.edu/abs/2014A&A...570A.114M} {570, A114}

\bibitem[\protect\citeauthoryear{{Marshall}, {Maddison}, {Thilliez},
  {Matthews}, {Wilner}, {Greaves}  \& {Holland}}{{Marshall}
  et~al.}{2017}]{2017Marshall}
{Marshall} J.~P.,  {Maddison} S.~T.,  {Thilliez} E.,  {Matthews} B.~C.,
  {Wilner} D.~J.,  {Greaves} J.~S.,   {Holland} W.~S.,  2017, \mn@doi [\mnras]
  {10.1093/mnras/stx645}, \href
  {https://ui.adsabs.harvard.edu/abs/2017MNRAS.468.2719M} {468, 2719}

\bibitem[\protect\citeauthoryear{{Marshall}, {Milli}, {Choquet}, {del Burgo},
  {Kennedy}, {Matr{\`a}}, {Ertel}  \& {Boccaletti}}{{Marshall}
  et~al.}{2018}]{2018Marshall}
{Marshall} J.~P.,  {Milli} J.,  {Choquet} {\'E}.,  {del Burgo} C.,  {Kennedy}
  G.~M.,  {Matr{\`a}} L.,  {Ertel} S.,   {Boccaletti} A.,  2018, \mn@doi [\apj]
  {10.3847/1538-4357/aaec6a}, \href
  {https://ui.adsabs.harvard.edu/abs/2018ApJ...869...10M} {869, 10}

\bibitem[\protect\citeauthoryear{{Marshall}, {Wang}, {Kennedy}, {Zeegers}  \&
  {Scicluna}}{{Marshall} et~al.}{2021}]{2021Marshall}
{Marshall} J.~P.,  {Wang} L.,  {Kennedy} G.~M.,  {Zeegers} S.~T.,   {Scicluna}
  P.,  2021, \mn@doi [\mnras] {10.1093/mnras/staa3917}, \href
  {https://ui.adsabs.harvard.edu/abs/2021MNRAS.501.6168M} {501, 6168}

\bibitem[\protect\citeauthoryear{{Matr{\`a}}, {Marino}, {Kennedy}, {Wyatt},
  {{\"O}berg}  \& {Wilner}}{{Matr{\`a}} et~al.}{2018}]{2018Matra}
{Matr{\`a}} L.,  {Marino} S.,  {Kennedy} G.~M.,  {Wyatt} M.~C.,  {{\"O}berg}
  K.~I.,   {Wilner} D.~J.,  2018, \mn@doi [\apj] {10.3847/1538-4357/aabcc4},
  \href {https://ui.adsabs.harvard.edu/abs/2018ApJ...859...72M} {859, 72}

\bibitem[\protect\citeauthoryear{{Matr{\`a}}, {{\"O}berg}, {Wilner}, {Olofsson}
   \& {Bayo}}{{Matr{\`a}} et~al.}{2019a}]{2019aMatra}
{Matr{\`a}} L.,  {{\"O}berg} K.~I.,  {Wilner} D.~J.,  {Olofsson} J.,   {Bayo}
  A.,  2019a, \mn@doi [\aj] {10.3847/1538-3881/aaff5b}, \href
  {https://ui.adsabs.harvard.edu/abs/2019AJ....157..117M} {157, 117}

\bibitem[\protect\citeauthoryear{{Matr{\`a}}, {Wyatt}, {Wilner}, {Dent},
  {Marino}, {Kennedy}  \& {Milli}}{{Matr{\`a}} et~al.}{2019b}]{2019bMatra}
{Matr{\`a}} L.,  {Wyatt} M.~C.,  {Wilner} D.~J.,  {Dent} W.~R.~F.,  {Marino}
  S.,  {Kennedy} G.~M.,   {Milli} J.,  2019b, \mn@doi [\aj]
  {10.3847/1538-3881/ab06c0}, \href
  {https://ui.adsabs.harvard.edu/abs/2019AJ....157..135M} {157, 135}

\bibitem[\protect\citeauthoryear{{Michel}, {van der Marel}  \&
  {Matthews}}{{Michel} et~al.}{2021}]{2021Michel}
{Michel} A.,  {van der Marel} N.,   {Matthews} B.~C.,  2021, \mn@doi [\apj]
  {10.3847/1538-4357/ac1bbb}, \href
  {https://ui.adsabs.harvard.edu/abs/2021ApJ...921...72M} {921, 72}

\bibitem[\protect\citeauthoryear{{Millar-Blanchaer} et~al.,}{{Millar-Blanchaer}
  et~al.}{2015}]{Millar-Blanchaer2015}
{Millar-Blanchaer} M.~A.,  et~al., 2015, \mn@doi [\apj]
  {10.1088/0004-637X/811/1/18}, 811, 18

\bibitem[\protect\citeauthoryear{{Milli}, {Mouillet}, {Lagrange}, {Boccaletti},
  {Mawet}, {Chauvin}  \& {Bonnefoy}}{{Milli} et~al.}{2012}]{Milli2012}
{Milli} J.,  {Mouillet} D.,  {Lagrange} A.-M.,  {Boccaletti} A.,  {Mawet} D.,
  {Chauvin} G.,   {Bonnefoy} M.,  2012, \mn@doi [\aap]
  {10.1051/0004-6361/201219687}, \href
  {http://cdsads.u-strasbg.fr/abs/2012A%26A...545A.111M} {545, A111}

\bibitem[\protect\citeauthoryear{{Mo{\'o}r} et~al.,}{{Mo{\'o}r}
  et~al.}{2011}]{2011Moor}
{Mo{\'o}r} A.,  et~al., 2011, \mn@doi [\apjs] {10.1088/0067-0049/193/1/4},
  \href {https://ui.adsabs.harvard.edu/abs/2011ApJS..193....4M} {193, 4}

\bibitem[\protect\citeauthoryear{{Mo{\'o}r} et~al.,}{{Mo{\'o}r}
  et~al.}{2015}]{2015Moor}
{Mo{\'o}r} A.,  et~al., 2015, \mn@doi [\mnras] {10.1093/mnras/stu2442}, \href
  {https://ui.adsabs.harvard.edu/abs/2015MNRAS.447..577M} {447, 577}

\bibitem[\protect\citeauthoryear{{Mo{\'o}r} et~al.,}{{Mo{\'o}r}
  et~al.}{2017}]{2017Moor}
{Mo{\'o}r} A.,  et~al., 2017, \mn@doi [\apj] {10.3847/1538-4357/aa8e4e}, \href
  {https://ui.adsabs.harvard.edu/abs/2017ApJ...849..123M} {849, 123}

\bibitem[\protect\citeauthoryear{{Mulders}, {Pascucci}, {Ciesla}  \&
  {Fernandes}}{{Mulders} et~al.}{2021}]{2021Mulders}
{Mulders} G.~D.,  {Pascucci} I.,  {Ciesla} F.~J.,   {Fernandes} R.~B.,  2021,
  \mn@doi [\apj] {10.3847/1538-4357/ac178e}, \href
  {https://ui.adsabs.harvard.edu/abs/2021ApJ...920...66M} {920, 66}

\bibitem[\protect\citeauthoryear{{Mustill} \& {Wyatt}}{{Mustill} \&
  {Wyatt}}{2009}]{2009MustillWyatt}
{Mustill} A.~J.,  {Wyatt} M.~C.,  2009, \mn@doi [\mnras]
  {10.1111/j.1365-2966.2009.15360.x}, \href
  {https://ui.adsabs.harvard.edu/abs/2009MNRAS.399.1403M} {399, 1403}

\bibitem[\protect\citeauthoryear{{Najita}, {Kenyon}  \& {Bromley}}{{Najita}
  et~al.}{2022}]{2022Najita}
{Najita} J.~R.,  {Kenyon} S.~J.,   {Bromley} B.~C.,  2022, \mn@doi [\apj]
  {10.3847/1538-4357/ac37b6}, \href
  {https://ui.adsabs.harvard.edu/abs/2022ApJ...925...45N} {925, 45}

\bibitem[\protect\citeauthoryear{{Norfolk} et~al.,}{{Norfolk}
  et~al.}{2021}]{2021Norfolk}
{Norfolk} B.~J.,  et~al., 2021, \mn@doi [\mnras] {10.1093/mnras/stab1901},
  \href {https://ui.adsabs.harvard.edu/abs/2021MNRAS.507.3139N} {507, 3139}

\bibitem[\protect\citeauthoryear{{Pan} \& {Schlichting}}{{Pan} \&
  {Schlichting}}{2012}]{2012Pan}
{Pan} M.,  {Schlichting} H.~E.,  2012, \mn@doi [\apj]
  {10.1088/0004-637X/747/2/113}, \href
  {https://ui.adsabs.harvard.edu/abs/2012ApJ...747..113P} {747, 113}

\bibitem[\protect\citeauthoryear{{Pawellek}, {Wyatt}, {Matr{\`a}}, {Kennedy}
  \& {Yelverton6}}{{Pawellek} et~al.}{2021}]{2021Pawellek}
{Pawellek} N.,  {Wyatt} M.,  {Matr{\`a}} L.,  {Kennedy} G.,   {Yelverton6} B.,
  2021, \mn@doi [\mnras] {10.1093/mnras/stab269}, \href
  {https://ui.adsabs.harvard.edu/abs/2021MNRAS.502.5390P} {502, 5390}

\bibitem[\protect\citeauthoryear{{Pearce} et~al.,}{{Pearce}
  et~al.}{2022}]{2022Pearce}
{Pearce} T.~D.,  et~al., 2022, \mn@doi [\aap] {10.1051/0004-6361/202142720},
  \href {https://ui.adsabs.harvard.edu/abs/2022A&A...659A.135P} {659, A135}

\bibitem[\protect\citeauthoryear{{Pollack}, {Hubickyj}, {Bodenheimer},
  {Lissauer}, {Podolak}  \& {Greenzweig}}{{Pollack} et~al.}{1996}]{1996Pollack}
{Pollack} J.~B.,  {Hubickyj} O.,  {Bodenheimer} P.,  {Lissauer} J.~J.,
  {Podolak} M.,   {Greenzweig} Y.,  1996, \mn@doi [\icarus]
  {10.1006/icar.1996.0190}, \href
  {https://ui.adsabs.harvard.edu/abs/1996Icar..124...62P} {124, 62}

\bibitem[\protect\citeauthoryear{{Ruane} et~al.,}{{Ruane}
  et~al.}{2019}]{Ruane2019}
{Ruane} G.,  et~al., 2019, \mn@doi [\aj] {10.3847/1538-3881/aafee2}, \href
  {https://ui.adsabs.harvard.edu/abs/2019AJ....157..118R} {157, 118}

\bibitem[\protect\citeauthoryear{{STScI Development Team}}{{STScI Development
  Team}}{2013}]{Pysynphot2013}
{STScI Development Team} 2013, {pysynphot: Synthetic photometry software
  package}, Astrophysics Source Code Library, record ascl:1303.023 (\mn@eprint
  {ascl} {1303.023})

\bibitem[\protect\citeauthoryear{{Schlichting}}{{Schlichting}}{2014}]{2014Schlichting}
{Schlichting} H.~E.,  2014, \mn@doi [\apjl] {10.1088/2041-8205/795/1/L15},
  \href {https://ui.adsabs.harvard.edu/abs/2014ApJ...795L..15S} {795, L15}

\bibitem[\protect\citeauthoryear{{Schneider} et~al.,}{{Schneider}
  et~al.}{2014}]{2014Schneider}
{Schneider} G.,  et~al., 2014, \mn@doi [\aj] {10.1088/0004-6256/148/4/59},
  \href {https://ui.adsabs.harvard.edu/abs/2014AJ....148...59S} {148, 59}

\bibitem[\protect\citeauthoryear{{Skrutskie} et~al.,}{{Skrutskie}
  et~al.}{2006}]{2006Skrutskie}
{Skrutskie} M.~F.,  et~al., 2006, \mn@doi [\aj] {10.1086/498708}, \href
  {https://ui.adsabs.harvard.edu/abs/2006AJ....131.1163S} {131, 1163}

\bibitem[\protect\citeauthoryear{{Soummer}, {Pueyo}  \& {Larkin}}{{Soummer}
  et~al.}{2012}]{Soummer2012}
{Soummer} R.,  {Pueyo} L.,   {Larkin} J.,  2012, \mn@doi [\apjl]
  {10.1088/2041-8205/755/2/L28}, 755, L28

\bibitem[\protect\citeauthoryear{{Soummer} et~al.,}{{Soummer}
  et~al.}{2014}]{Soummer2014}
{Soummer} R.,  et~al., 2014, \mn@doi [\apjl] {10.1088/2041-8205/786/2/L23},
  786, L23

\bibitem[\protect\citeauthoryear{{Su{\'a}rez Mascare{\~n}o}
  et~al.,}{{Su{\'a}rez Mascare{\~n}o} et~al.}{2021}]{2022SuarezMascareno}
{Su{\'a}rez Mascare{\~n}o} A.,  et~al., 2021, \mn@doi [Nature Astronomy]
  {10.1038/s41550-021-01533-7}, \href
  {https://ui.adsabs.harvard.edu/abs/2022NatAs...6..232S} {6, 232}

\bibitem[\protect\citeauthoryear{{Tazzari}, {Beaujean}  \& {Testi}}{{Tazzari}
  et~al.}{2018}]{2018Tazzari}
{Tazzari} M.,  {Beaujean} F.,   {Testi} L.,  2018, \mn@doi [\mnras]
  {10.1093/mnras/sty409}, \href
  {http://adsabs.harvard.edu/abs/2018MNRAS.476.4527T} {476, 4527}

\bibitem[\protect\citeauthoryear{{Th{\'e}bault}}{{Th{\'e}bault}}{2009}]{Thebault2009}
{Th{\'e}bault} P.,  2009, \mn@doi [\aap] {10.1051/0004-6361/200912396}, 505,
  1269

\bibitem[\protect\citeauthoryear{{Tychoniec} et~al.,}{{Tychoniec}
  et~al.}{2020}]{2020Tychoniec}
{Tychoniec} {\L}.,  et~al., 2020, \mn@doi [\aap] {10.1051/0004-6361/202037851},
  \href {https://ui.adsabs.harvard.edu/abs/2020A&A...640A..19T} {640, A19}

\bibitem[\protect\citeauthoryear{{Vican}, {Schneider}, {Bryden}, {Melis},
  {Zuckerman}, {Rhee}  \& {Song}}{{Vican} et~al.}{2016}]{2016Vican}
{Vican} L.,  {Schneider} A.,  {Bryden} G.,  {Melis} C.,  {Zuckerman} B.,
  {Rhee} J.,   {Song} I.,  2016, \mn@doi [\apj] {10.3847/1538-4357/833/2/263},
  \href {https://ui.adsabs.harvard.edu/abs/2016ApJ...833..263V} {833, 263}

\bibitem[\protect\citeauthoryear{Virtanen et~al.,}{Virtanen
  et~al.}{2020}]{2020Virtanen}
Virtanen P.,  et~al., 2020, \mn@doi [Nature Methods]
  {10.1038/s41592-019-0686-2}, \href {https://rdcu.be/b08Wh} {17, 261}

\bibitem[\protect\citeauthoryear{{Wahhaj} et~al.,}{{Wahhaj}
  et~al.}{2016}]{Wahhaj2016}
{Wahhaj} Z.,  et~al., 2016, \mn@doi [\aap] {10.1051/0004-6361/201321887}, \href
  {http://cdsads.u-strasbg.fr/abs/2016arXiv161105866W} {596, L4}

\bibitem[\protect\citeauthoryear{{Wright} et~al.,}{{Wright}
  et~al.}{2010}]{2010Wright}
{Wright} E.~L.,  et~al., 2010, \mn@doi [\aj] {10.1088/0004-6256/140/6/1868},
  \href {https://ui.adsabs.harvard.edu/abs/2010AJ....140.1868W} {140, 1868}

\bibitem[\protect\citeauthoryear{{Zhang} et~al.,}{{Zhang}
  et~al.}{2018}]{2018Zhang}
{Zhang} S.,  et~al., 2018, \mn@doi [\apjl] {10.3847/2041-8213/aaf744}, \href
  {https://ui.adsabs.harvard.edu/abs/2018ApJ...869L..47Z} {869, L47}

\bibitem[\protect\citeauthoryear{{del Burgo} \& {Allende Prieto}}{{del Burgo}
  \& {Allende Prieto}}{2016}]{2016delBurgo}
{del Burgo} C.,  {Allende Prieto} C.,  2016, \mn@doi [\mnras]
  {10.1093/mnras/stw2005}, \href
  {https://ui.adsabs.harvard.edu/abs/2016MNRAS.463.1400D} {463, 1400}

\bibitem[\protect\citeauthoryear{{del Burgo} \& {Allende Prieto}}{{del Burgo}
  \& {Allende Prieto}}{2018}]{2018delBurgo}
{del Burgo} C.,  {Allende Prieto} C.,  2018, \mn@doi [\mnras]
  {10.1093/mnras/sty1371}, \href
  {https://ui.adsabs.harvard.edu/abs/2018MNRAS.479.1953D} {479, 1953}

\bibitem[\protect\citeauthoryear{{van der Marel} \& {Mulders}}{{van der Marel}
  \& {Mulders}}{2021}]{2021vanderMarel}
{van der Marel} N.,  {Mulders} G.~D.,  2021, \mn@doi [\aj]
  {10.3847/1538-3881/ac0255}, \href
  {https://ui.adsabs.harvard.edu/abs/2021AJ....162...28V} {162, 28}

\bibitem[\protect\citeauthoryear{{van der Marel}, {Dong}, {di Francesco},
  {Williams}  \& {Tobin}}{{van der Marel} et~al.}{2019}]{2019vanderMarel}
{van der Marel} N.,  {Dong} R.,  {di Francesco} J.,  {Williams} J.~P.,
  {Tobin} J.,  2019, \mn@doi [\apj] {10.3847/1538-4357/aafd31}, \href
  {https://ui.adsabs.harvard.edu/abs/2019ApJ...872..112V} {872, 112}

\makeatother
\end{thebibliography}

%\begin{thebibliography}{99}
%\end{thebibliography}

%%%%%%%%%%%%%%%%%%%%%%%%%%%%%%%%%%%%%%%%%%%%%%%%%%

%%%%%%%%%%%%%%%%% APPENDICES %%%%%%%%%%%%%%%%%%%%%

%\appendix

%%%%%%%%%%%%%%%%%%%%%%%%%%%%%%%%%%%%%%%%%%%%%%%%%%
\appendix

\section{Posterior probability distributions of the disc modelling}

\begin{figure*}
    \centering
    \includegraphics[width=\textwidth]{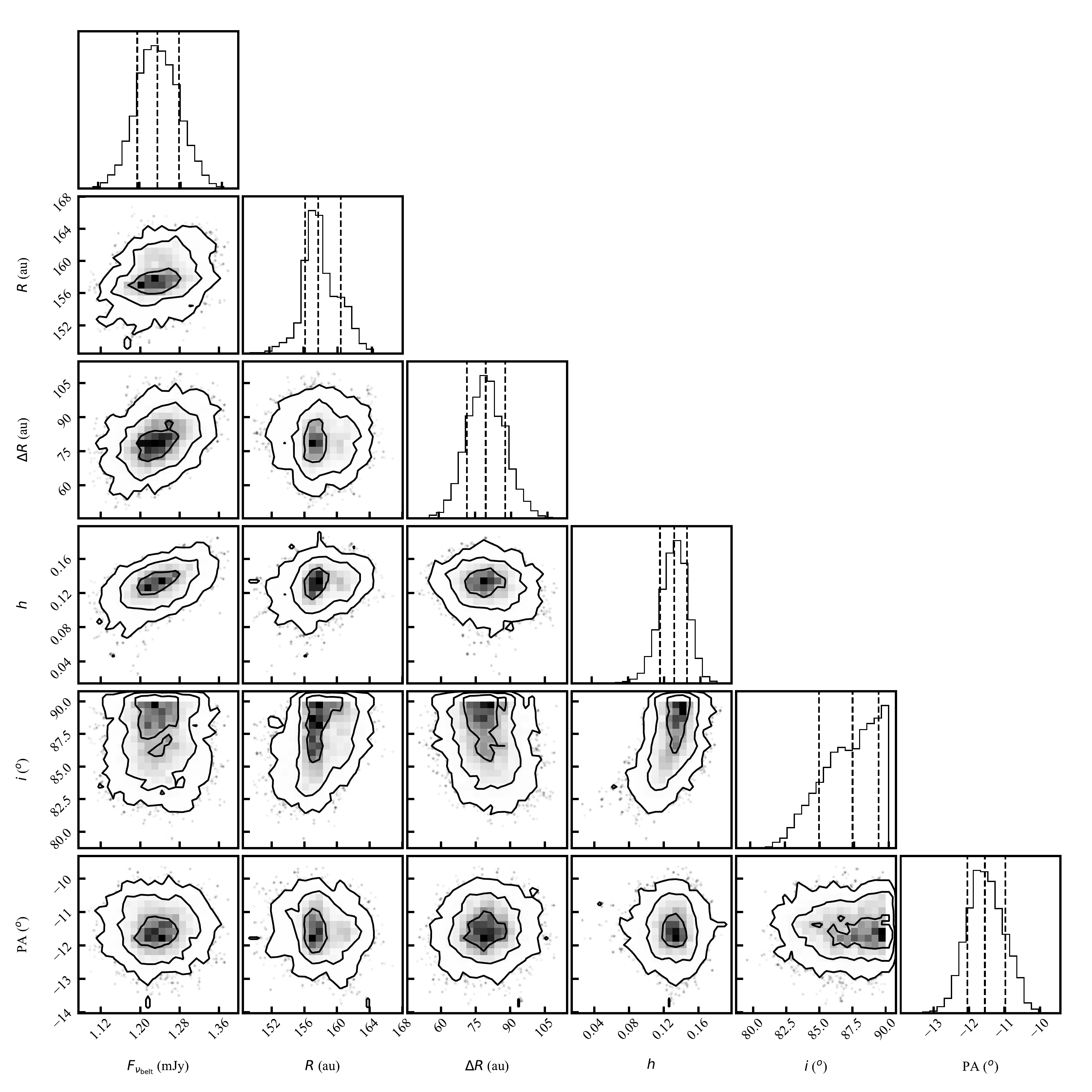}
    \caption{Corner plot showing the posterior probability distributions for the {\sc emcee} runs used to identify the maximum amplitude probability model and calculate the uncertainties. The posteriors are mono-modal and well behaved except for the inclination which runs up against the edge of the allowed parameter space (90$\degr$). A small degeneracy between the scale height and the inclination is seen in the 2D distribution of these posteriors, expressed as a long tail in their combined distribution toward lower scale heights and inclinations.}
\end{figure*}

% Don't change these lines
\bsp	% typesetting comment
\label{lastpage}
\end{document}